\documentclass[aps,prd,preprint,nofootinbib,amsmath,amssymb,superscriptaddress,floatfix]{revtex4-2}

\usepackage{graphicx}

\makeatletter
\@addtoreset{equation}{section}
\makeatother

\usepackage{subcaption}

\usepackage{hyperref}
\hypersetup{colorlinks=true, linkcolor=blue, citecolor=magenta, urlcolor=blue}

\usepackage{mathtools}

\usepackage{ragged2e}

\begin{document}

\title{Modular Flow and Holographic Timelike Entanglement Entropy in  AdS and Flat Holography}

\author{Mehdi Hakami Shalamzari}
\email[Corresponding author: ]{mehdi.hakami@gmail.com, m.hakami@scu.ac.ir}
\affiliation{Physics Department, Faculty of Science, Shahid Chamran University of Ahvaz, Ahvaz 61357-43135, Iran}

\author{Zahra Paygozar}
\email{zahrapaygozar79@gmail.com}
\affiliation{Physics Department, Faculty of Science, Shahid Chamran University of Ahvaz, Ahvaz 61357-43135, Iran}

\author{Mahdi Kord Zangeneh}
\email{mkzangeneh@scu.ac.ir}
\affiliation{Physics Department, Faculty of Science, Shahid Chamran University of Ahvaz, Ahvaz 61357-43135, Iran}

\begin{abstract}
	Timelike entanglement entropy provides a useful probe of holographic theories, yet an explicit bulk realization of the modular flow underlying its timelike extremal curve construction has so far been lacking. We construct modular flows associated with spacelike and timelike intervals in AdS$_{3}$/CFT$_{2}$ and extend this framework to flat$_{3}$/CCFT$_{2}$. Beginning with the boundary  symmetry generators, we explicitly derive the boundary modular flows and extend them into the bulk, providing a systematic procedure that relates bulk modular flows to the extremal curves and holographic (timelike) entanglement entropy.
For spacelike intervals, the resulting bulk modular flows determine the extremal curves, whose integrated line elements reproduce the standard holographic entanglement entropy. We then generalize this approach to timelike intervals, explicitly constructing the modular flows whose fixed points reproduce the extremal spacelike and timelike curves, which give the holographic timelike entanglement entropy. To the best of our knowledge, this is the first explicit construction of the modular flow associated with a timelike extremal curve for a timelike interval.
The same procedure is then carried over to flat$_{3}$/CCFT$_{2}$, which gives rise to the holographic (timelike) entanglement entropy. Finally, we show that the flat space results obtained in this work can be recovered from their AdS$_3$ counterparts  in the flat space limit.
\end{abstract}

\maketitle
\thispagestyle{empty}
\newpage

\tableofcontents

	\section{Introduction}

	Gauge/gravity duality \cite{Maldacena_1999,witten1998antisitterspaceholography,Gubser_1998} establishes a profound correspondence between quantum information in the boundary theory and geometric structures in the bulk, with entanglement entropy providing a key manifestation of this relation. This duality provides a dictionary through which intricate quantum field theoretic quantities acquire geometric and gravitational interpretations in the bulk. Gauge/gravity duality has also found applications beyond high-energy physics, including in condensed matter systems and strongly coupled quantum systems \cite{stangier2026holographicsuperconductivitycriticalfermi,liu2024quantumcorrectionsholographicstrange,Ahn_2025,Cinti_2025,Tai_2026,steiner2026holographicquantumcodestrapped,Green_2013}.

The most prominent example of gauge/gravity duality is the celebrated AdS/CFT correspondence \cite{Maldacena_1999}. The AdS/CFT correspondence relates gravity in a $(d+1)$-dimensional asymptotically Anti-de Sitter (AdS) spacetime to a $d$-dimensional conformal field theory (CFT)  living on its boundary. Among the most significant applications of this correspondence is the holographic description of entanglement entropy. Direct evaluation of the entanglement entropy of a generic subsystem in a CFT is typically difficult. Ryu and Takayanagi  nonetheless proposed an elegant geometric prescription for this quantity \cite{Ryu_2006,Ryu_2007}, later generalized covariantly \cite{Hubeny_2007}. According to this proposal, the entanglement entropy of a boundary subsystem $A$ is determined by the area of a codimension-two extremal surface $\gamma_{A}$ in the asymptotically AdS bulk, homologous to $A$ and anchored on the boundary of $A$, divided by $4G$,
\begin{equation}
S_A = \dfrac{\mathrm{Area}(\gamma_A)}{4G_N},
\end{equation}
with $G_{N}$ Newton's gravitational constant. A derivation of this prescription directly from two-dimensional CFT data
has also been given recently, in which the dual AdS$_3$ geometry emerges
from an ensemble of large-$c$ CFTs instead of being assumed
\cite{Bao2504,Geng2505}.

This relation carries broader significance: the geometry of the bulk spacetime directly reflects the entanglement structure of the boundary theory, suggesting that spacetime geometry may emerge from the underlying structure of quantum entanglement \cite{VanRaamsdonk:2010pw,Swingle_2012}. Holographic entanglement entropy therefore plays a central role in the AdS/CFT correspondence, encoding deep aspects of the bulk geometry. While these developments were established primarily in the context of the AdS/CFT correspondence, a natural question is whether analogous ideas persist beyond asymptotically AdS geometries, and in particular for asymptotically flat spacetimes.

In recent years, considerable effort has been devoted to extending holography beyond asymptotically AdS spacetimes, including proposals such as flat/CCFT, dS/CFT \cite{strominger2001dscft,witten2001quantumgravitydesitter}, Kerr/CFT  \cite{guica2009kerrcft}, and celestial holography~\cite{pasterski2017flatspaceamplitudes,pasterski2021celestialholography,pasterski2021lecturescelestialamplitudes}. Among these, we focus on the flat/CCFT correspondence, which relates asymptotically flat spacetimes in $(d+1)$-dimensions to Carrollian conformal field theories (CCFTs) living on their $d$-dimensional null boundaries \cite{PhysRevLett.105.171601,Bagchi_2012}. These field theories arise as Carrollian (ultrarelativistic) limits of ordinary conformal field theories, obtained by sending the speed of light to zero \cite{Levy-Leblond:1965dsc,Duval_20140,Duval_2014}. CCFTs in two and three dimensions possess infinite-dimensional symmetry algebras that closely mirror the asymptotic symmetry algebras of asymptotically flat spacetimes at null infinity, namely the BMS symmetries in three and four dimensions \cite{Barnich_20100,Barnich:2006av,Barnich_2010}. These algebras are isomorphic, lending strong support to the flat/CCFT conjecture.

As in AdS/CFT, observables of the boundary theory are expected to admit holographic counterparts in the gravitational dual theory. Entanglement entropy for CCFT subsystems is similarly expected to be tightly constrained by the infinite-dimensional symmetries of the theory. Holographic proposals for entanglement entropy in asymptotically flat spacetimes have been developed. For two-dimensional CCFTs, the computation reduces to identifying extremal surfaces in the asymptotically flat bulk geometry. Unlike their AdS counterparts, however, these surfaces are not directly anchored to the endpoints of the boundary interval; instead, they connect to the corresponding subsystem at null infinity through two null rays~\cite{Jiang_2017}.

Conventional entanglement entropy is intrinsically associated with spacelike subsystems, and initial studies of entangling behavior in CFTs and CCFTs were correspondingly limited to this regime. The extension to timelike subsystems via the notion of pseudo entropy was made in \cite{Doi:2022iyj,Doi_20231}.
Pseudo entropy \cite{Nakata_2021,Murciano_2022,Guo_2022,Guo_2023,he2023pseudoentropyprimaryoperators} generalizes entanglement entropy by replacing the reduced density matrix with a transition matrix built from two distinct quantum states. Timelike entanglement entropy emerges as a particular realization of pseudo entropy, obtained by analytically continuing a spacelike subsystem to a timelike one. The resulting quantity is generically complex: its real part is associated with the area of spacelike extremal surfaces, while its imaginary part is tied to timelike surfaces in the bulk geometry.
In AdS$_{3}$/CFT$_{2}$, the holographic dual of timelike entanglement entropy consists of two spacelike extremal curves anchored at the endpoints of the timelike boundary interval and joined by a timelike curve in the bulk, forming a natural geometric extension of holographic entanglement entropy to timelike regions \cite{Doi:2022iyj,Doi_20231}. The flat limit of these curves is well defined and gives rise to the holographic timelike entanglement entropy in the context of flat$_{3}$/CCFT$_{2}$ holography \cite{Fareghbal_2026}; our result for its imaginary part is compared with theirs in Sec.~\ref{Timelike Entanglement Entropy in Flat Space from Modular Flow}.
It should be noted that the imaginary part of timelike entanglement entropy may encode information about the emergence of the time coordinate, extending the idea that spatial geometry emerges from ordinary entanglement entropy \cite{Doi:2022iyj,Doi_20231}. 

The work of \cite{Doi:2022iyj,Doi_20231} has stimulated a growing
literature on timelike subsystems. One line of research asks what the
imaginary part means geometrically: the geometric interpretation of
timelike entanglement entropy and the extraction of temporal
entanglement from holographic entanglement entropy are addressed in
\cite{Heller2408,Heller2507}, while \cite{Roy2507} follows how the
entropy interpolates between spacelike and timelike configurations, and
\cite{Roy2508} extends the holographic picture to higher dimensions.
A second line modifies the gravitational side, with gravitational
anomalies \cite{Anom2504} and higher-curvature corrections
\cite{HC2509}, and studies the first law of entanglement
\cite{Sun2511}. Further work relates pseudo entropy to the phase
structure of entanglement \cite{Kanda2311} and to ergodicity and
spectral chaos \cite{Chaos2601}. In flat space, the limit of the
holographic curves and the resulting Carrollian pseudo entropy are
treated in \cite{Fareghbal:2024lqa,Fareghbal_2026}, and the relation to
the inner horizon in \cite{Wen2025}. Pseudo entropy has also been studied in field-theoretic settings, such as
free gauge theory, where its relation to real-time evolution through
analytic continuation was examined \cite{Mukherjee2205}, and timelike
entanglement entropy has been considered in non-relativistic holography,
in the Lifshitz background, where both the real and imaginary parts
receive corrections \cite{Jena2410}. Related constructions of modular
flow in other geometries include celestial conformal field theory
\cite{Ghodrati2605}, and timelike intervals have been used to study
string scattering amplitudes in confining backgrounds
\cite{Ghodrati2307}. For Carrollian and Galilean theories, boosted
intervals and the corresponding limits were used in the study of
symmetry-resolved entanglement \cite{Banerjee2404}.

Computing HEE and HTEE requires solving a generically nontrivial extremization problem to identify the correct bulk surfaces. An alternative approach to obtaining the extremal curves, and hence computing HEE and HTEE, relies on the modular flow. For a given subsystem, the modular flow generates a one-parameter family of transformations describing the evolution of operators with respect to the modular Hamiltonian \cite{Casini:2011kv,Haag:1992hx}.  In the bulk, this flow is constructed from Killing vectors of the background geometry. Within the holographic framework, extremal surfaces coincide with the set of fixed points of the bulk modular flow~\cite{Apolo_2020}. In particular, the extremal surface remains invariant under the bulk modular flow. For a single interval in the vacuum of a CFT$_{2}$, where the modular flow is geometric, this property allows the extremal curve to be read off from symmetry, without solving the extremization problem. This method, as we will show, extends naturally to flat space holography as well.

The flat-space limit is where the modular flow approach has a clear
advantage. The extremal curves of Ryu and Takayanagi do not have a
well-defined flat limit in general. In~\cite{Fareghbal:2020dtq} a new
vector field was introduced in AdS$_3$, which is orthogonal to the bulk
modular flow of the original interval and vanishes on a curve whose
flat limit is well defined. After a Rindler transformation the bulk
modular flow becomes the Killing vector normal to the outer horizon of
the BTZ black hole, while the new vector becomes the Killing vector
normal to the inner horizon. The flat limit of this new vector
reproduces the bulk modular flow used for BMS-invariant field theories,
and the resulting curves give the holographic entanglement entropy of
CCFT$_2$ \cite{Fareghbal:2020dtq}.

The connection with timelike intervals was made in
\cite{Fareghbal:2024lqa}. There it was shown that the curves of
\cite{Fareghbal:2020dtq} are the spacelike branches of the extremal
curves of the pseudo entropy of a timelike interval in AdS$_3$, and
that if their length is computed from the radial coordinate at which
the retarded time is extremal, the flat limit reproduces the
entanglement entropy of CCFT$_2$ with $c_L=0$ and $c_M=3/G_N$. The same
holds in dS$_3$. In \cite{Fareghbal_2026} the timelike curve was shown
to have a well-defined flat limit as well, so that the lengths of the
spacelike and timelike curves give the real and imaginary parts of a
pseudo entropy in CCFT$_2$; in this sense the entanglement entropy of a
CCFT is itself a pseudo entropy. In these works the curves are obtained
from the extremal curves of AdS$_3$ and their flat limits. What has been
missing is a derivation of both curves, spacelike and timelike, from the
symmetries of the boundary theory, in the way the flow of
\cite{Fareghbal:2020dtq} does for the spacelike branches, together with
a flow whose fixed points are the timelike curve.

In this paper we fill this gap. We construct the boundary modular flows
of spacelike and timelike intervals in CFT$_2$ and CCFT$_2$ from the
symmetry generators, extend them into the bulk, and show that their
fixed points are the curves of
\cite{Fareghbal:2020dtq,Fareghbal:2024lqa,Fareghbal_2026}. The flow
$\xi_{st}$ below is the vector of \cite{Fareghbal:2020dtq}; it also
appears as the modular momentum flow in the study of the inner horizon
in timelike entanglement \cite{Wen2025}. Our new ingredient is a second
flow, $\xi_{tt}$, whose fixed points are the timelike curve, and its
flat limit. The geometric meaning of these flows in the Rindler picture,
including the position of the timelike curve between the two horizons
of the BTZ black hole and in the flat-space cosmology, is the subject of
the companion paper \cite{shalamzari2026holographicpseudoentropyrindler}.

Modular flow establishes a link between quantum information and spacetime geometry, simultaneously encoding the quantum entanglement structure of the boundary theory and determining geometric objects in the bulk. Thus, modular flow is a useful tool for probing holographic dualities in both asymptotically AdS and flat spacetimes. In this work, we construct the boundary and bulk modular flows and use them to investigate the holographic entanglement entropy associated with both spacelike and timelike subsystems.

The remainder of this paper is organized as follows. In Sec. \ref{Preliminaries}, we review the basic concepts and conventions used throughout this work. In  Sec. \ref{Entanglement Entropy from Modular Flow}, we treat spacelike intervals in AdS$_{3}$/CFT$_{2}$, constructing the modular flow on the boundary and extending it into the bulk, and using the fixed point structure of this flow to determine the extremal surfaces and the corresponding holographic entanglement entropy. In Sec. \ref{Timelike Entanglement Entropy from Modular Flow}, we generalize this analysis to timelike intervals, following the same logic to obtain the timelike entanglement entropy. In Sec. \ref{Timelike Entanglement Entropy in Flat Space from Modular Flow}, we extend the construction to the flat$_{3}$/CCFT$_{2}$ correspondence, treating both spacelike and timelike intervals in Carrollian conformal field theories. Finally, Sec. \ref{Conclusion} summarizes our results and discusses future directions.

\section{Preliminaries}
\label{Preliminaries}

Here we review the background needed for the rest of the paper. Entanglement entropy and its generalization to pseudo entropy are discussed first, with particular attention to timelike subsystems. This is followed by a summary of the holographic realization of entanglement entropy, in both the AdS$_{3}$/CFT$_{2}$ and  flat$_{3}$/CCFT$_{2}$
correspondences. Finally, we close this section with the modular Hamiltonian and modular flow, whose geometric interpretation in the bulk underlies the constructions of Sections \ref{Entanglement Entropy from Modular Flow}--\ref{Timelike Entanglement Entropy in Flat Space from Modular Flow}.

\subsection{Entanglement Entropy and Pseudo Entropy}
Consider a quantum system partitioned into a subsystem $A$ and its complement $B$. The Hilbert space of the total system admits a tensor product decomposition,
\begin{equation}
	\mathcal{H}=\mathcal{H}_{A}\otimes \mathcal{H}_{B},
\end{equation}
where the subsystem $A$ is generally taken to be an arbitrary spatial region, though for simplicity we restrict to a fixed time slice. This decomposition provides the basic framework for defining entanglement entropy through the reduced density matrix of the subsystem $A$.

If the total system is described by a pure state $|\psi\rangle$, its density matrix is given by
\begin{equation}
	\rho = |\psi\rangle\langle\psi|.
\end{equation}
The reduced density matrix of the subsystem $A$ is obtained by tracing out the degrees of freedom of its complement $B$,
\begin{equation}
	\rho_A=\mathrm{Tr}_{B}(\rho).
	\label{26}
\end{equation}
The entanglement entropy of the subsystem $A$ is then defined as the von Neumann entropy of the reduced density matrix,
\begin{equation}
	S_A=-\mathrm{Tr}\left(\rho_{A}\log\rho_{A}\right).
\end{equation}
This quantity measures the entanglement between the subsystem $A$ and its complement $B$.

While entanglement entropy is defined for a single quantum state, pseudo entropy provides a natural extension to a pair of quantum states. Consider two states $|\psi\rangle$ and $|\varphi\rangle$, which are not necessarily identical. The corresponding transition matrix is defined as \cite{Nakata_2021}
\begin{equation}
	\tau=\frac{|\psi\rangle\langle\varphi|}{\langle\varphi|\psi\rangle}.
\end{equation}
Similar to the construction of the reduced density matrix, the reduced transition matrix associated with the subsystem $A$ is obtained by tracing out the degrees of freedom of its complement $B$,
\begin{equation}
	\tau_A=\mathrm{Tr}_{B}(\tau).
\end{equation}
The pseudo entropy is then defined as the von Neumann entropy of the reduced transition matrix,
\begin{equation}
	S_A^{(p)}=-\mathrm{Tr}\left(\tau_A\log\tau_A\right).
\end{equation}
Unlike entanglement entropy, which is a real quantity, pseudo entropy captures the quantum correlation between two distinct states and can, in general, take complex values. Importantly, when the two states coincide, namely $|\psi\rangle=|\varphi\rangle$, the transition matrix reduces to the standard density matrix, and the pseudo entropy naturally reproduces the entanglement entropy.

A natural generalization arises when the subsystem $A$ is chosen to be a timelike interval on the boundary rather than a spacelike one. In this case, an entanglement entropy associated with the subsystem can still be defined. However, a crucial difference emerges: the corresponding reduced density matrix is no longer Hermitian. Consequently, the resulting entanglement entropy is not necessarily real and can generally acquire an imaginary contribution. This quantity is referred to as timelike entanglement entropy (TEE) \cite{Doi_20231}.

The non-Hermitian nature of the reduced density matrix suggests that timelike entanglement entropy should be understood within a broader framework. In fact, it can be naturally formulated in terms of a transition matrix connecting two quantum states, in close analogy with the definition of pseudo entropy \cite{Doi:2022iyj,Doi_20231}. From this perspective, timelike entanglement entropy can be viewed as a particular realization of pseudo entropy, where the non-Hermitian structure of the transition matrix leads to the characteristic complex behavior of the entropy.

\subsection{AdS and Flat Holography and Holographic Entanglement Entropy}
The AdS$_{3}$/CFT$_{2}$ correspondence provides a well-established realization of holography, in which a three-dimensional asymptotically AdS spacetime is dual to a two-dimensional conformal field theory living on its boundary \cite{Maldacena_1999,witten1998antisitterspaceholography}. The asymptotic symmetry algebra of AdS$_{3}$ consists of two copies of the Witt algebra, generated by
\begin{equation}
	\begin{split}
		\left[L_{m},L_{n}\right] &= \left(m-n\right) L_{m+n}, \\[0.8ex]
		\left[\bar{L}_{m},\bar{L}_{n}\right] &= \left(m-n\right) \bar{L}_{m+n}, \\[0.8ex]
		\left[L_{m},\bar{L}_{n}\right] &= 0, \qquad m,n \in \mathbb{Z},
	\end{split}
\end{equation}
which is exactly the conformal algebra of CFT in two dimensions. The algebra of conserved charges is centrally extended and the corresponding central charges are $c=\bar{c}=3\ell/\left(2G\right)$~\cite{Brown:1986nw}.

In the holographic description, the entanglement entropy of a spatial subregion $A$ in the boundary CFT$_2$ is determined by the length of a codimension-two extremal curve in AdS$_3$ anchored at the endpoints of $A$.

Another important realization of holography arises in asymptotically flat spacetimes through the Flat$_3$/CCFT$_2$ correspondence \cite{Barnich_2010,PhysRevLett.105.171601,Bagchi_2012}. In the bulk side, this duality can be understood as the flat space limit of the AdS$_3$/CFT$_2$ correspondence, obtained by taking the AdS radius $\ell$ to infinity. On the boundary side, this procedure corresponds to an ultra-relativistic (İnönü–Wigner) contraction of the conformal symmetry, obtained by sending the speed of light to zero. As a result, the two copies of the Witt algebra characterizing the CFT$_2$ are contracted into the infinite-dimensional Carroll algebra in two dimensions \cite{PhysRevLett.105.171601}, which is isomorphic to the $\mathfrak{bms_3}$ algebra, the asymptotic symmetry algebra of asymptotic flat space-times in three dimensions. This algebra takes the following form
\begin{equation}
	\begin{split}
		\left[L_{m},L_{n}\right] &= \left(m-n\right) L_{m+n}, \\[0.8ex]
		\left[L_{m},M_{n}\right] &= \left(m-n\right) M_{m+n}, \\[0.8ex]
		\left[M_{m},M_{n}\right] &= 0, \qquad m,n \in \mathbb{Z},
	\end{split}
\end{equation}
where $M_{m}$ and $L_{m}$  are, respectively, the generators of supertranslations and superrotations. The algebra of conserved charges is centrally extended with central charges $c_L=0$ and $c_M=3/G_N$.

The holographic entanglement entropy in this case is the length of an extremal spacelike curve between two null rays coming from null infinity.

\subsection{Modular Flow and Modular Hamiltonian} 
For a quantum system partitioned into two subsystems, $A$ and $B$, the reduced density matrix associated with the subsystem $A$ is obtained by tracing out the degrees of freedom of subsystem $B$, as given in \eqref{26}. The reduced density matrix allows one to define a Hermitian operator, known as the modular Hamiltonian, through
\begin{equation}
	\rho_A=e^{-H_A}.
\end{equation}

It is important to emphasize that, in general, the modular Hamiltonian $H_A$ is non-local. In other words, it cannot always be expressed solely in terms of quantum fields supported within the subsystem $A$. Nevertheless, there exist special cases in which the modular Hamiltonian admits a local expression. A notable example is provided by the vacuum state of a quantum field theory restricted to a half-space in Minkowski spacetime. In this case, the modular Hamiltonian generates Lorentz boosts that preserve the corresponding Rindler wedge \cite{Bisognano:1976za}. Another important example is the vacuum state of a conformal field theory in which the subsystem $A$ is restricted to a ball-shaped region, for which the modular Hamiltonian also admits a local expression \cite{Casini:2011kv}.

This observation naturally leads to the concept of modular flow. While the modular Hamiltonian is a Hermitian operator characterizing the reduced density matrix, the modular flow describes the one-parameter family of transformations generated by it \cite{Haag:1992hx}. The modular Hamiltonian is, up to a constant, the conserved charge associated with this flow. The modular evolution of an operator $\mathcal{O}$ is defined as
\begin{equation}
	\mathcal{O}(s)=e^{iH_As}\mathcal{O}e^{-iH_As},
\end{equation}
where $s$ denotes the modular parameter. Equivalently, the modular flow can be expressed as
\begin{equation}
	\frac{d}{ds}\mathcal{O}(s)=i[H_A,\mathcal{O}(s)].
\end{equation}

In the context of holography, the boundary modular flow of a highly symmetric state and subregion extends into the bulk as a Killing vector field that preserves the entanglement wedge. This vector field vanishes on the corresponding extremal surface. The extension relies on the locality of $H_A$, which holds for the vacuum and an interval (or a ball), but not in general.

\section{Entanglement Entropy from Modular Flow }
\label{Entanglement Entropy from Modular Flow}
In this section, we construct the boundary and bulk modular flow associated with a spacelike interval in  AdS$_{3}$/CFT$_2$. We begin by reviewing the symmetry generators of the corresponding field theory on the boundary. Since the boundary theory is described by a two-dimensional conformal field theory, the relevant symmetry algebra consists of two copies of the Witt algebra, whose generators act on the boundary and admit well-defined extensions into the bulk. To construct the boundary modular flow we need only globally well-defined CFT symmetry generators. We write their explicit form and their bulk counterpart in Poincar\'e coordinates and use them to construct the corresponding boundary and bulk modular flow for a spacelike interval. 

The AdS$_{3}$ geometry in Poincar\'e coordinates is described by the metric
\begin{equation}
	ds^{2}=\dfrac{\ell^{2}}{z^{2}}(dx^{2}+dz^{2}-dt^{2})
	\label{eq:poincare_metric}
\end{equation}
where $\ell$ denotes the AdS radius, and the conformal boundary is located at $z=0$.
As mentioned above, the symmetry generators of CFT$_{2}$ are described by two copies of the Witt algebra. The corresponding globally well-defined boundary symmetry generators~\footnote{Throughout this paper, superscripts $b$ and $B$ denote boundary and bulk quantities, respectively, including generators and modular flows.} can be written as

\begin{equation}
\begin{split}
	L^{b}_{-1} &= - \dfrac{1}{2}\left(\partial_{t} + \partial_{x}\right), \\
	L^{b}_{0} &= - \dfrac{t+x}{2} \partial_{t} -\dfrac{t+x}{2} \partial_{x}, \\
	L^{b}_{1} &= - \dfrac{t^2 + x^2 + 2 t x}{2} \partial_{t} - \dfrac{t^2 + x^2 + 2 t x}{2} \partial_{x},
\end{split}
\label{40}
\end{equation} 
with
\begin{equation}
\begin{split}
	\bar{L}^{b}_{-1} &= - \dfrac{1}{2}\left(\partial_{t} - \partial_{x}\right), \\
	\bar{L}^{b}_{0} &= - \dfrac{t - x}{2} \partial_{t} + \dfrac{t - x}{2} \partial_{x}, \\
	\bar{L}^{b}_{1} &= - \dfrac{t^2 + x^2 - 2 t x}{2} \partial_{t} + \dfrac{t^2 + x^2 - 2 t x}{2} \partial_{x}.
\end{split}
\end{equation}
Extending these generators into the bulk leads to vector fields preserving the AdS$_{3}$ asymptotic structure. Their explicit bulk realization is given by
\begin{equation}
\begin{split}
	L^{B}_{-1} &= - \dfrac{1}{2}\left(\partial_{t} + \partial_{x}\right), \\
	L^{B}_{0} &= - \dfrac{t+x}{2} \partial_{t} -\dfrac{t+x}{2} \partial_{x} -\dfrac{z}{2} \partial_{z}, \\
	L^{B}_{1} &= - \dfrac{t^2 + x^2 + 2 t x + z^2}{2} \partial_{t} - \dfrac{t^2 + x^2 + 2 t x - z^2}{2} \partial_{x} - z (t + x)\,\partial_{z},
\end{split}
\end{equation} 
with
\begin{equation}
\begin{split}
	\bar{L}^{B}_{-1} &= - \dfrac{1}{2}\left(\partial_{t} - \partial_{x}\right), \\
	\bar{L}^{B}_{0} &= - \dfrac{t - x}{2} \partial_{t} + \dfrac{t - x}{2} \partial_{x} -\dfrac{z}{2} \partial_{z}, \\
	\bar{L}^{B}_{1} &= - \dfrac{t^2 + x^2 - 2 t x + z^2}{2} \partial_{t} + \dfrac{t^2 + x^2 - 2 t x - z^2}{2} \partial_{x} - z (t - x)\,\partial_{z}.
\end{split}
\label{41}
\end{equation}
Now, we consider the case in which the interval on the boundary is spacelike,
\begin{equation}
	-R<x<R ,\qquad t=0, 
	\label{61}
\end{equation}
where $R$ denotes the half-length of the boundary interval.

For a spacelike boundary interval, we consider boundary and bulk modular flows $\xi^b_{ss}$ and $\xi^B_{ss}$\footnote{The subscript $ss$ refers to a spacelike boundary interval with a spacelike bulk extremal curve. The subscripts $st$ and $tt$, introduced in Sec.~\ref{Timelike Entanglement Entropy from Modular Flow}, denote timelike boundary intervals with spacelike and timelike bulk extremal curves, respectively.} constructed from the CFT$_2$ symmetry generators and the corresponding AdS$_3$ Killing vectors
respectively. 
Explicitly, the standard boundary modular flow is given by \cite{Das:2023yyl}
\begin{equation}
	\begin{split}
		\xi^{b}_{ss} = \dfrac{\pi}{R} \Bigl[ \left(L^{b}_{1} + \bar{L}^{b}_{1}\right) - R^2 \left(L^{b}_{-1} + \bar{L}^{b}_{-1}\right) \Bigr] 
		= -\dfrac{2 \pi t x}{R}\, \partial_{x} + \dfrac{\pi}{R}\left(R^{2}-x^{2}-t^{2}\right) \partial_{t}.
	\end{split}
\end{equation}
The corresponding bulk modular flow is obtained by extending the boundary generators into the bulk and repeating the boundary construction with these extended generators,
\begin{equation}
	\begin{split}
		\xi^{B}_{ss} = \dfrac{\pi}{R} \Bigl[ \left(L^{B}_{1} + \bar{L}^{B}_{1}\right) - R^2 \left(L^{B}_{-1} + \bar{L}^{B}_{-1}\right) \Bigr] 
		= -\dfrac{2 \pi t}{R}\left(x \partial_{x} + z \partial_{z}\right) + \dfrac{\pi}{R}\left(R^{2}-x^{2}-t^{2}-z^{2}\right) \partial_{t}.
	\end{split}
	\label{36}
\end{equation}
The boundary modular flow is manifestly recovered from its bulk counterparts in the limit z$\rightarrow$0.

Since the extremal surface corresponding to the boundary subregion $A$ coincides with the  set of fixed points of the bulk modular flow, imposing the vanishing of the bulk modular flow vector in \eqref{36} yields
\begin{equation}
	x^{2}+z^{2}=R^2,\qquad t=0,
\end{equation}
which is precisely the bulk extremal curve anchored to the corresponding spacelike boundary interval that is shown as the green curve in Fig.~\ref{fig:1}.

To generalize our treatment to an arbitrary spacelike boundary interval, we perform a Lorentz boost on the boundary coordinates and construct the corresponding boosted modular flows. The Lorentz boost is given by
\begin{equation}
	t^{\prime} = t \cosh \eta + x \sinh \eta, \qquad x^{\prime} = t \sinh \eta + x \cosh \eta,
	\label{7}
\end{equation}
where $\eta$ is the rapidity parameter. The boosted boundary interval is specified by 
\begin{equation}
	-\dfrac{l_{x}}{2} < x^{\prime} < \dfrac{l_{x}}{2}, \qquad -\dfrac{l_{t}}{2} < t^{\prime} < \dfrac{l_{t}}{2}.
\end{equation}
The rapidity parameter associated with the boosted interval is uniquely fixed by
\begin{equation}
	\eta = \log \sqrt{\dfrac{l_{x} + l_{t}}{l_{x} - l_{t}}},
\end{equation}
while the invariant length of the interval is
\begin{equation}
	R^{2} = \dfrac{l_{x}^{2}-l_{t}^{2}}{4}.
\end{equation}
The condition $l_{x} > l_{t}$ ensures that the interval is spacelike.

Using these relations, the boosted boundary vector field can be written as
\begin{equation}
	\begin{split}
		\xi_{ss}^{b} = \dfrac{\pi \left(-l_{t}^{2} l_{x}+l_{x}^3+8 l_{t} x^{\prime} t^{\prime} - 4 l_{x}\left(x^{\prime2} + t^{\prime2}\right)\right)}{2 \left(l_{x}^2-l_{t}^2\right)} \partial_{t^{\prime}} 
	 +  \dfrac{\pi \left(l_{x}^{2} l_{t}-l_{t}^3-8 l_{x} x^{\prime} t^{\prime} + 4 l_{t}\left(x^{\prime2} + t^{\prime2}\right)\right)}{2 \left(l_{x}^2-l_{t}^2\right)} \partial_{x^{\prime}}.
	\end{split}
\end{equation}
By extending this vector field into the bulk, its bulk counterpart can be rewritten in the following form:
\begin{equation}
	\begin{split}
		\xi_{ss}^{B} &= \dfrac{\pi \left(-l_{t}^{2} l_{x}+l_{x}^3+8 l_{t} x^{\prime} t^{\prime} - 4 l_{x}\left(x^{\prime2} + t^{\prime2} + z^{2}\right)\right)}{2 \left(l_{x}^2-l_{t}^2\right)} \partial_{t^{\prime}} \\{}& \quad + \dfrac{\pi \left(l_{x}^{2} l_{t}-l_{t}^3-8 l_{x} x^{\prime} t^{\prime} + 4 l_{t}\left(x^{\prime2} + t^{\prime2} - z^{2}\right)\right)}{2 \left(l_{x}^2-l_{t}^2\right)} \partial_{x^{\prime}}\\{}& \quad
		 + \dfrac{4 \pi \left( l_{t} x^{\prime}-l_{x} t^{\prime}\right) z}{l_{x}^{2}-l_{t}^2} \partial_{z}.
	\end{split}
	\label{1}
\end{equation}

Following the same procedure, the corresponding extremal curve is obtained by imposing the vanishing of the bulk modular flow components. For the bulk modular flow \eqref{1}, this yields
\begin{equation}
	x^{\prime2} = \dfrac{l_{x}^{2}(l_{x}^2-l_{t}^{2}-4z^{2})}{4(l_{x}^{2}-l_{t}^{2})} , \qquad t^{\prime2} = \dfrac{l_{t}^{2}(l_{x}^2-l_{t}^{2}-4z^{2})}{4(l_{x}^{2}-l_{t}^{2})}.
\end{equation}

The flat space limit $\ell\rightarrow\infty$ is not well defined in the Poincar\'e coordinates. We therefore use the following transformation  to introduce the global BMS coordinates~\cite{Fareghbal:2024lqa}:
\begin{equation}
	\begin{split}
		t &= \dfrac{2 \ell \left(r \sin\left(\dfrac{u}{\ell}\right) - \ell \cos\left(\dfrac{u}{\ell}\right)\right)}{\Delta}, \\ x &= \dfrac{2 \ell r \sin\left(\phi\right)}{\Delta}, \\ z &= \dfrac{2 \ell^{2}}{\Delta}, \\ \Delta &= \ell \sin\left( \dfrac{u}{\ell}\right) + r \cos \left(\dfrac{u}{\ell}\right) + r \cos \left(\phi\right).
		\label{8}
	\end{split}
\end{equation}
Under this transformation, the Poincar\'e metric \eqref{eq:poincare_metric} takes the form
\begin{equation}
	ds^{2} = - \left(1 + \dfrac{r^{2}}{\ell^{2}} \right) du^{2} - 2 du dr + r^{2} d\phi^{2}.
	\label{17}
\end{equation}

We consider a spacelike boundary interval in the global BMS coordinates for the case of the CFT$_2$ vacuum state on the cylinder, parameterized by
\begin{equation}
	-\dfrac{l_{u}}{2} < u < \dfrac{l_{u}}{2}, \qquad -\dfrac{l_{\phi}}{2} < \phi < \dfrac{l_{\phi}}{2}. 
\end{equation}
The corresponding interval on the Poincar\'e boundary is characterized by the lengths 
\begin{equation}
	\begin{split}
	l_{x} = \dfrac{4 \ell \sin \left(\dfrac{l_{\phi}}{2}\right)}{\cos \left(\dfrac{l_{\phi}}{2}\right) + \cos \left(\dfrac{l_{u}}{2 \ell}\right)},\qquad	l_{t} = \dfrac{4 \ell \sin \left(\dfrac{l_{u}}{2 \ell}\right)}{\cos \left(\dfrac{l_{\phi}}{2}\right) + \cos \left(\dfrac{l_{u}}{2 \ell}\right)}.
\end{split}
\end{equation}
The requirement that the interval be spacelike, $l_{x} >l_{t} $, translates into $ \sin \left(\dfrac{l_{\phi}}{2}\right) > \sin \left(\dfrac{l_{u}}{2 \ell}\right) $.

In the global BMS coordinates, the boosted boundary modular flow takes the form
\begin{widetext}
	\begin{footnotesize}
		\begin{equation}
			\begin{split}
				\xi_{ss}^{b} &= \dfrac{2 \pi \ell \left(- 2 \cos \left(\phi \right) \sin \left(\dfrac{l_{\phi}}{2}\right) \cos \left(\dfrac{u}{\ell}\right) \cos \left(\dfrac{l_{u}}{2 \ell}\right) - 2 \sin \left(\phi \right) \cos \left(\dfrac{l_{\phi}}{2}\right) \sin\left( \dfrac{u}{\ell} \right) \sin \left(\dfrac{l_{u}}{2 \ell} \right) + \sin \left(l_{\phi}\right)\right)}{\cos \left(l_{\phi}\right) - \cos \left(\dfrac{l_{u}}{\ell}\right)} \partial_{u} \\{}& \quad + \dfrac{2 \pi \left(2 \cos \left(\phi\right) \cos \left(\dfrac{l_{\phi}}{2}\right) \sin \left(\dfrac{l_{u}}{2 \ell}\right)  \cos \left(\dfrac{u}{\ell}\right)  + 2 \sin \left(\phi\right) \sin \left(\dfrac{l_{\phi}}{2}\right) \cos \left(\dfrac{l_{u}}{2 \ell}\right)  \sin \left(\dfrac{u}{\ell}\right)  -  \sin \left(\dfrac{l_{u}}{\ell}\right)\right)}{\cos \left(l_{\phi}\right) - \cos \left(\dfrac{l_{u}}{\ell}\right)} \partial_{\phi}.
			\end{split}
		\end{equation}
	\end{footnotesize}
\end{widetext}
The corresponding boosted bulk modular flow is given~by~\cite{Fareghbal:2020dtq}
\begin{widetext}
	\begin{scriptsize}
		\begin{equation}
			\begin{split}
				\xi_{ss}^{B} &= \dfrac{2 \pi \ell \left(- 2 \cos \left(\phi \right) \sin \left(\dfrac{l_{\phi}}{2}\right) \cos \left(\dfrac{u}{\ell}\right) \cos \left(\dfrac{l_{u}}{2 \ell}\right) - 2 \sin \left(\phi \right) \cos \left(\dfrac{l_{\phi}}{2}\right) \sin\left( \dfrac{u}{\ell} \right) \sin \left(\dfrac{l_{u}}{2 \ell} \right) + \sin \left(l_{\phi}\right)\right)}{\cos \left(l_{\phi}\right) - \cos \left(\dfrac{l_{u}}{\ell}\right)} \partial_{u} \\[0.7ex]{}& \quad + \dfrac{4 \pi \left(\cos \left(\phi\right) \sin \left(\dfrac{l_{\phi}}{2}\right) \cos \left(\dfrac{l_{u}}{2 \ell}\right)\left(\ell \cos \left(\dfrac{u}{\ell}\right) - r \sin \left(\dfrac{u}{\ell}\right)\right) + \sin \left(\phi\right) \cos \left(\dfrac{l_{\phi}}{2}\right) \sin \left(\dfrac{l_{u}}{2 \ell}\right) \left(r \cos \left(\dfrac{u}{\ell}\right) + \ell \sin \left(\dfrac{u}{\ell}\right)\right)\right)}{\cos \left(l_{\phi}\right) - \cos \left(\dfrac{l_{u}}{\ell}\right)} \partial_{r} \\[0.7ex]{}& \quad + \dfrac{2 \pi \left(2 \cos \left(\phi\right) \cos \left(\dfrac{l_{\phi}}{2}\right) \sin \left(\dfrac{l_{u}}{2 \ell}\right) \left(r \cos \left(\dfrac{u}{\ell}\right) + \ell \sin \left(\dfrac{u}{\ell}\right)\right) + 2 \sin \left(\phi\right) \sin \left(\dfrac{l_{\phi}}{2}\right) \cos \left(\dfrac{l_{u}}{2 \ell}\right) \left(r \sin \left(\dfrac{u}{\ell}\right) - \ell \cos \left(\dfrac{u}{\ell}\right)\right) - r \sin \left(\dfrac{l_{u}}{\ell}\right)\right)}{r\left(\cos \left(l_{\phi}\right) - \cos \left(\dfrac{l_{u}}{\ell}\right)\right)} \partial_{\phi}.
			\end{split}
			\label{38}
		\end{equation}
	\end{scriptsize}
\end{widetext}

We now want to determine the extremal curve associated with the flow. The extremal curve corresponding to the modular flow in~\eqref{38} is given by the following relations 
\begin{equation}
	\begin{split}
	r = \dfrac{\ell \cos \left(\dfrac{l_{\phi}}{2}\right) \sin \left(\dfrac{u}{\ell}\right)}{\cos \left(\dfrac{l_{u}}{2 \ell}\right) \cos \left(\phi\right) - \cos \left(\dfrac{l_{\phi}}{2}\right) \cos \left(\dfrac{u}{\ell}\right)}, \\
	-r \sin \left(\phi\right) \sin \left(\dfrac{l_{u}}{2 \ell}\right) = \sin \left(\dfrac{l_{\phi}}{2}\right) (\ell \cos \left(\dfrac{u}{\ell}\right) - r \sin \left(\dfrac{u}{\ell}\right)),
	\end{split}
	\label{19}
\end{equation}
the resulting relations are in agreement with those obtained in \cite{Fareghbal:2024lqa}.

For computing the holographic entanglement entropy of the spacelike interval, we need the line element of the corresponding extremal curve. Hence, substituting the extremal curve \eqref{19} into the metric \eqref{17} yields the following line element
\begin{equation}
	\begin{split}
		ds_{\xi_{ss}} =  \dfrac{\ell \left(\cos \left(\dfrac{l_{u}}{\ell}\right) - \cos \left(l_{\phi}\right)\right) r \: dr}{\sqrt{\left(r^2 \left(\cos \left(\dfrac{l_{u}}{\ell}\right) - \cos \left(l_{\phi}\right)\right) - \ell^{2} \cos \left(l_{\phi}\right)\right)^2 - \ell^4}}
	\end{split}
	\label{39}
\end{equation}
Evaluating the length in \eqref{39} requires determining the integration range of the radial coordinate $r$. With the help of equations \eqref{19}, we obtain
\begin{equation}
	r^2 = \dfrac{\ell^2 \sin^{2} \left(l_{\phi}\right)}{\left(\cos \left(l_{\phi}\right) - \cos \left(2 \phi\right)\right) \left(\cos \left(l_{\phi}\right) - \cos \left(\dfrac{l_{u}}{\ell}\right)\right)}.
\end{equation}
Imposing $\dfrac{\partial r}{\partial \phi}=0$ yields the minimum value of the radial coordinate,

\begin{equation}
	r_{min} = \dfrac{\ell \cos \left(\dfrac{l_{\phi}}{2}\right)}{\sqrt{\sin^{2} \left(\dfrac{l_{\phi}}{2}\right) - \sin^{2}\left(\dfrac{l_{u}}{2 \ell}\right)}}
\end{equation}
The upper limit corresponds to the AdS boundary, $r_{\max}\rightarrow\infty$. To regularize the AdS boundary divergence, we introduce the cutoff  $r_{\varepsilon} = \ell^2/\varepsilon $.
Thus,
\begin{equation}
	\begin{split}
	L_{ss} &= 2 \int_{r_{min}}^{r_{\varepsilon}} ds_{\xi_{ss}} = \ell \log \left[\dfrac{2 \ell^2}{\varepsilon^2} \left(\cos \left(\dfrac{l_{u}}{\ell}\right) -\cos \left(l_{\phi}\right)\right)\right] \\[0.4ex]{}&
	=\ell \log\left[\dfrac{4 \ell^2}{\varepsilon^2} \left(\sin \left(\dfrac{l_{\phi}}{2} + \dfrac{l_{u}}{2 \ell}\right) \sin \left(\dfrac{l_{\phi}}{2} - \dfrac{l_{u}}{2 \ell}\right)\right)\right].
		\end{split}
\end{equation}
We can  compute the entanglement entropy associated with the spacelike interval, which takes the form
\begin{equation}
	\begin{split}
	S_{ss} &= \dfrac{L_{ss}}{4 G}  = \dfrac{\ell}{4 G} \log\left[\dfrac{4 \ell^2}{\varepsilon^2} \left(\sin \left(\dfrac{l_{\phi}}{2} + \dfrac{l_{u}}{2 \ell}\right) \sin \left(\dfrac{l_{\phi}}{2} - \dfrac{l_{u}}{2 \ell}\right)\right)\right] \\[0.4ex]{}&
	= \dfrac{c}{6} \log\left[\dfrac{4 \ell^2}{\varepsilon^2} \left(\sin \left(\dfrac{l_{\phi}}{2} + \dfrac{l_{u}}{2 \ell}\right) \sin \left(\dfrac{l_{\phi}}{2} - \dfrac{l_{u}}{2 \ell}\right)\right)\right],
	\label{20}
		\end{split}
\end{equation}
where we have used the Brown--Henneaux relation $c=3\ell / \left(2G\right)$. A more familiar form of the entanglement entropy, corresponding to the standard CFT result on a cylinder \cite{Doi_20231}, can be obtained by substituting the relations $ \Delta t = l_{u}$ ,  $ \Delta \phi = \ell l_{\phi}$, and  $ \mathfrak{R} = 2 \pi \ell $ into \eqref{20}. This yields
\begin{equation}
	\begin{split}
		S_{ss} = \dfrac{c}{6} \log \Biggl[ \dfrac{\mathfrak{R}^2}{\pi^2 \varepsilon^2} \sin \left(\dfrac{\pi}{\mathfrak{R}}\left(\Delta \phi+ \Delta t\right)\right) \sin \left(\dfrac{\pi}{\mathfrak{R}} \left(\Delta \phi - \Delta t\right) \right) \Biggr].
	\end{split}
\end{equation}

In summary, we constructed the boundary and bulk modular flows associated with a spacelike interval in AdS$_{3}$/CFT$_{2}$, first in a t=0 time slice and then for a generic spacelike subsystem using a boundary Lorentz boost.  We then expressed the boosted flows in global BMS coordinates, which will be used also in taking the flat space limit in Sec.~\ref{Timelike Entanglement Entropy in Flat Space from Modular Flow}. By identifying the extremal curve with the fixed points of bulk modular flow, we obtained the holographic entanglement entropy and found it in agreement with the corresponding CFT result on the cylinder.

\section{Timelike Entanglement Entropy from Modular Flow}
\label{Timelike Entanglement Entropy from Modular Flow}

In the previous section, we considered spacelike intervals. We now turn to timelike intervals. As reviewed in Sec.~\ref{Preliminaries}, the  entanglement entropy of a timelike subsystem is generically complex, its real and imaginary parts encoding the lengths of distinct bulk curves. Here we construct the corresponding boundary and bulk modular flows, using their fixed-point structure to determine both contributions. The real part, treated in Sec.~\ref{sec:real part}, comes from the disconnected spacelike branches anchored at the interval endpoints. The timelike curve connecting them gives rise to  the imaginary part, discussed in Sec.~\ref{sec:imaginary_part}.

\subsection{Real Part}
\label{sec:real part}
\begin{figure}
	\centering
	\includegraphics[width=0.6\textwidth]{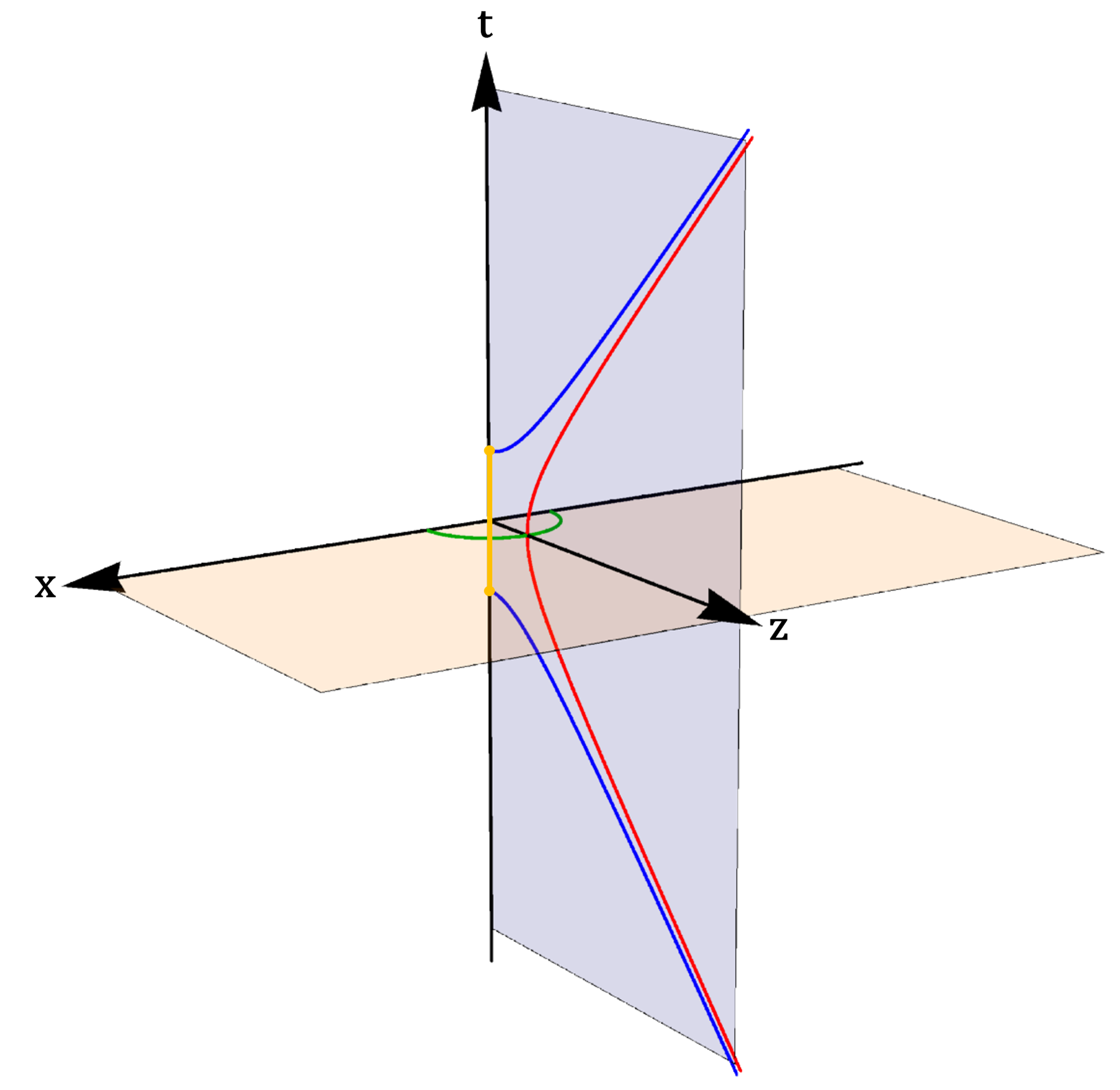}
	\caption{ \justifying Extremal curves associated with a timelike interval in the Poincar\'e patch before the boost transformation. The boundary timelike interval is shown in yellow. The blue and red curves correspond to the disconnected spacelike extremal curves and the timelike extremal curve, respectively, while the green curve represents the extremal curve associated with the vector field $\xi^B_{ss}$, which is the corresponding bulk modular flow of the spacelike interval \eqref{61}.}
	\label{fig:1}
\end{figure}

For convenience, let us consider a pure timelike interval in a CFT$_{2}$ specified by
\begin{equation}
	-T < t < T, \qquad x = 0,
	\label{9}
\end{equation}
where $T$ is an arbitrary positive constant. The corresponding dual bulk geometry is described by the AdS$_{3}$ metric in Poincar\'e coordinates, given by \eqref{eq:poincare_metric}.
Following the construction of~\cite{Das:2023yyl}, based on the CFT$_{2}$
symmetry generators, the boundary modular flow associated with this interval is given by
\begin{equation}
	\begin{split}
		\xi^{b}_{st} = \dfrac{\pi}{T} \Bigl[ \left(L^{b}_{1} - \bar{L}^{b}_{1}\right) - T^2 \left(L^{b}_{-1} - \bar{L}^{b}_{-1}\right) \Bigr] 
		= -\dfrac{2 \pi t x}{T}\, \partial_{t} + \dfrac{\pi}{T}\left(T^{2}-x^{2}-t^{2}\right) \partial_{x},
	\end{split}
\end{equation}
Upon extending this boundary vector fields into the bulk, we obtain the corresponding bulk modular flow as follows 
\begin{equation}
\begin{split}
	\xi^{B}_{st} = \dfrac{\pi}{T} \Bigl[ \left(L^{B}_{1} - \bar{L}^{B}_{1}\right) - T^2 \left(L^{B}_{-1} - \bar{L}^{B}_{-1}\right) \Bigr] 
	= -\dfrac{2 \pi x}{T}\left(t \partial_{t} + z \partial_{z}\right) + \dfrac{\pi}{T}\left(T^{2}-x^{2}-t^{2}+z^{2}\right) \partial_{x}.
\end{split}
\label{5}
\end{equation}
The extremal curve associated with the modular flow in \eqref{5} is 
\begin{equation}
	T^2 = t^2 - z^2, \qquad x = 0.
\end{equation}
This curve reproduces the disconnected spacelike extremal curve (see blue curves in Fig.~\ref{fig:1}) with two branches anchored at the endpoints of the interval \cite{Doi:2022iyj,Doi_20231}. 

Applying the boost transformation of \eqref{7} to the boundary coordinates, we obtain a more general boundary interval,
\begin{equation}
	-\dfrac{l_{x}}{2} < t^{\prime} < \dfrac{l_{x}}{2}, \quad \quad \quad -\dfrac{l_{t}}{2} < x^{\prime} < \dfrac{l_{t}}{2},
\end{equation}
with
\begin{equation}
	T^{2} = \dfrac{l_{x}^{2}-l_{t}^{2}}{4}, \qquad \eta = \log \sqrt{\dfrac{l_{x} + l_{t}}{l_{x} - l_{t}}}
\end{equation}
The timelike condition requires $l_x>l_t$. The corresponding boosted boundary vector field is given by
\begin{equation}
	\begin{split}
		\xi_{st}^{b} = \dfrac{\pi \left(-l_{x}^{2} l_{t}+l_{t}^3+8 l_{x} x^{\prime} t^{\prime} - 4 l_{t}\left(x^{\prime2} + t^{\prime2}\right)\right)}{2 \left(l_{t}^2-l_{x}^2\right)} \partial_{t^{\prime}} + \dfrac{\pi \left(l_{t}^{2} l_{x}-l_{x}^3-8 l_{t} x^{\prime} t^{\prime} + 4 l_{x}\left(x^{\prime2} + t^{\prime2}\right)\right)}{2 \left(l_{t}^2-l_{x}^2\right)} \partial_{x^{\prime}},
	\end{split}
\end{equation}
while the corresponding bulk vector field takes the form
\begin{equation}
	\begin{split}
		\xi_{st}^{B} &= \dfrac{\pi \left(-l_{x}^{2} l_{t}+l_{t}^3+8 l_{x} x^{\prime} t^{\prime} - 4 l_{t}\left(x^{\prime2} + t^{\prime2} + z^{2}\right)\right)}{2 \left(l_{t}^2-l_{x}^2\right)} \partial_{t^{\prime}} \\{}& \quad + \dfrac{\pi \left(l_{t}^{2} l_{x}-l_{x}^3-8 l_{t} x^{\prime} t^{\prime} + 4 l_{x}\left(x^{\prime2} + t^{\prime2} - z^{2}\right)\right)}{2 \left(l_{t}^2-l_{x}^2\right)} \partial_{x^{\prime}} \\{}& \quad - \dfrac{4 \pi \left(- l_{x} x^{\prime}+l_{t} t^{\prime}\right) z}{l_{t}^{2}-l_{x}^2} \partial_{z}.
	\end{split}
	\label{42}
\end{equation}
The extremal curve associated with the boosted modular flow in \eqref{42} is 
\begin{equation}
	x^{\prime2} = \dfrac{l_{t}^{2}(l_{x}^2-l_{t}^{2}+4z^{2})}{4(l_{x}^{2}-l_{t}^{2})} , \qquad t^{\prime2} = \dfrac{l_{x}^{2}(l_{x}^2-l_{t}^{2}+4z^{2})}{4(l_{x}^{2}-l_{t}^{2})} .
\end{equation}
This curve also appears in \cite{Fareghbal_2026}. 
By comparing  \eqref{1} and \eqref{42}, one finds that the two vector fields are exchanged under~\cite{Fareghbal:2024lqa}
\begin{equation}
	l_{x} \longleftrightarrow l_{t}. 
	\label{3}
\end{equation}
Moreover, these flows satisfy \cite{Fareghbal:2020dtq}
\begin{equation}
	\begin{split}
	\left(\xi_{ss}^{B}\right)^{\mu} \left(\xi_{st}^{B}\right)_{\mu} &= 0 \\
	\left(\xi_{ss}^{B}\right)^{\mu} \left(\xi_{ss}^{B}\right)_{\mu}  + \left(\xi_{st}^{B}\right)^{\mu} \left(\xi_{st}^{B}\right)_{\mu} &= 4 \pi^{2} \ell^2.
	\end{split}
	\label{4}
\end{equation}
These expressions show that $\xi_{ss}^{B}$ and $\xi_{st}^{B}$  are orthogonal and fulfill an interesting norm relation. 

We now transform to global BMS coordinates using the coordinate transformation in \eqref{8}. The corresponding timelike interval is given by 
\begin{equation}
	-\dfrac{\ell l_{\phi}}{2} < u < \dfrac{\ell l_{\phi}}{2},\qquad -\dfrac{l_{u}}{2 \ell} < \phi < \dfrac{l_{u}}{2 \ell}
	\label{10}
\end{equation}
which yields the following relations between the interval parameters $l_{x},l_{t}$ and $l_{u},l_{\phi}$:
\begin{equation}
	\begin{split}
	l_{x} = \dfrac{4 \ell \sin \dfrac{l_{\phi}}{2}}{\cos \dfrac{l_{\phi}}{2} + \cos \dfrac{l_{u}}{2 \ell}}, \qquad	l_{t} = \dfrac{4 \ell \sin \dfrac{l_{u}}{2 \ell}}{\cos \dfrac{l_{\phi}}{2} + \cos \dfrac{l_{u}}{2 \ell}}.
	\label{11}
	\end{split}
\end{equation}
Using the timelike condition $l_{x}>l_{t}$, we find $\sin \left(\dfrac{l_\phi}{2}\right)>\sin \left(\dfrac{l_u}{2\ell}\right)$.
The corresponding boundary vector field in global BMS coordinates is 
\begin{widetext}
		\begin{scriptsize}
		\begin{equation}
			\begin{split}
				\xi_{st}^{b} &= \dfrac{2 \pi \ell \left( 2 \cos \left(\phi \right) \sin \left(\dfrac{l_{u}}{2 \ell}\right) \cos \left(\dfrac{u}{\ell}\right) \cos \left(\dfrac{l_{\phi}}{2}\right) + 2 \sin \left(\phi \right) \cos \left(\dfrac{l_{u}}{2 \ell}\right) \sin\left( \dfrac{u}{\ell} \right) \sin \left(\dfrac{l_{\phi}}{2} \right) - \sin \left(\dfrac{l_{u}}{\ell}\right)\right)}{\cos \left(l_{\phi}\right) - \cos \left(\dfrac{l_{u}}{\ell}\right)} \partial_{u} \\{}& \quad + \dfrac{2 \pi \left(-2 \cos \left(\phi\right) \cos \left(\dfrac{l_{u}}{2 \ell}\right) \sin \left(\dfrac{l_{\phi}}{2}\right)  \cos \left(\dfrac{u}{\ell}\right)  - 2 \sin \left(\phi\right) \sin \left(\dfrac{l_{u}}{2 \ell}\right) \cos \left(\dfrac{l_{\phi}}{2}\right)  \sin \left(\dfrac{u}{\ell}\right) +  \sin \left(l_{\phi}\right)\right)}{\cos \left(l_{\phi}\right) - \cos \left(\dfrac{l_{u}}{\ell}\right)} \partial_{\phi},
			\end{split}
			\label{77}
		\end{equation}
			\end{scriptsize}
\end{widetext}
Its bulk extension is

\begin{widetext}
	\begin{scriptsize}
		\begin{equation}
			\begin{split}
				\xi_{st}^{B} &= \dfrac{2 \pi \ell \left( 2 \cos \left(\phi \right) \sin \left(\dfrac{l_{u}}{2 \ell}\right) \cos \left(\dfrac{u}{\ell}\right) \cos \left(\dfrac{l_{\phi}}{2}\right) + 2 \sin \left(\phi \right) \cos \left(\dfrac{l_{u}}{2 \ell}\right) \sin\left( \dfrac{u}{\ell} \right) \sin \left(\dfrac{l_{\phi}}{2} \right) - \sin \left(\dfrac{l_{u}}{\ell}\right)\right)}{\cos \left(l_{\phi}\right) - \cos \left(\dfrac{l_{u}}{\ell}\right)} \partial_{u} \\[0.7ex]{}& \quad - \dfrac{4 \pi \left(\cos \left(\phi\right) \sin \left(\dfrac{l_{u}}{2 \ell}\right) \cos \left(\dfrac{l_{\phi}}{2}\right)\left(\ell \cos \left(\dfrac{u}{\ell}\right) - r \sin \left(\dfrac{u}{\ell}\right)\right) + \sin \left(\phi\right) \cos \left(\dfrac{l_{u}}{2 \ell}\right) \sin \left(\dfrac{l_{\phi}}{2}\right) \left(r \cos \left(\dfrac{u}{\ell}\right) + \ell \sin \left(\dfrac{u}{\ell}\right)\right)\right)}{\cos \left(l_{\phi}\right) - \cos \left(\dfrac{l_{u}}{\ell}\right)} \partial_{r} \\[0.7ex]{}& \quad+ \dfrac{2 \pi \left(-2 \cos \left(\phi\right) \cos \left(\dfrac{l_{u}}{2 \ell}\right) \sin \left(\dfrac{l_{\phi}}{2}\right) \left(r \cos \left(\dfrac{u}{\ell}\right) + \ell \sin \left(\dfrac{u}{\ell}\right)\right) + 2 \sin \left(\phi\right) \sin \left(\dfrac{l_{u}}{2 \ell}\right) \cos \left(\dfrac{l_{\phi}}{2}\right) \left(\ell \cos \left(\dfrac{u}{\ell}\right) - r \sin \left(\dfrac{u}{\ell}\right)\right) + r \sin \left(l_{\phi}\right)\right)}{r\left(\cos \left(l_{\phi}\right) - \cos \left(\dfrac{l_{u}}{\ell}\right)\right)} \partial_{\phi},
			\end{split}
			\label{44}
		\end{equation}
	\end{scriptsize}
\end{widetext}

The modular flow in \eqref{44} yields the following extremal curve
\begin{equation}
	\begin{split}
	r = \dfrac{\ell \cos \left(\dfrac{l_{u}}{2 \ell}\right) \sin \left(\dfrac{u}{\ell}\right)}{\cos \left(\dfrac{l_{\phi}}{2}\right) \cos \left(\phi\right) - \cos \left(\dfrac{l_{u}}{2 \ell}\right) \cos \left(\dfrac{u}{\ell}\right)},\\
	-r \sin \left(\phi\right) \sin \left(\dfrac{l_{\phi}}{2}\right) = \sin \left(\dfrac{l_{u}}{2 \ell}\right) \left(\ell \cos \left(\dfrac{u}{\ell}\right) - r \sin \left(\dfrac{u}{\ell}\right)\right),
	\end{split}
	\label{16}
\end{equation}
reproducing the extremal curve obtained in \cite{Fareghbal_2026}. The two vector fields $\xi_{ss}^{B}$ and $\xi_{st}^{B}$ in \eqref{38} and \eqref{44} and thereby their associated extremal curves in \eqref{19} and \eqref{16} are related by the parameter transformation
\begin{equation}
	l_{u} \longrightarrow \ell l_{\phi} \qquad l_{\phi} \longrightarrow \dfrac{l_{u}}{\ell}.
\end{equation}
From equation \eqref{16}, we obtain
\begin{equation}
	r^2 = \dfrac{\ell^2 \sin^{2} \left(\dfrac{l_{u}}{\ell}\right)}{\left(\cos \left(\dfrac{l_{u}}{\ell}\right) - \cos \left(2 \phi\right)\right) \left(\cos \left(\dfrac{l_{u}}{\ell}\right) - \cos \left(l_{\phi}\right)\right)}.
\end{equation}
Minimizing $r^{2}$ with respect to $\phi$ gives 
\begin{equation}
	r_{min} = \dfrac{\ell \sin \left(\dfrac{l_{u}}{2 \ell}\right)}{\sqrt{\sin^{2} \left(\dfrac{l_{\phi}}{2}\right) - \sin^{2}\left(\dfrac{l_{u}}{2 \ell}\right)}}.
	\label{31}
\end{equation}

Using the corresponding extremal curve equations in \eqref{16} and the metric \eqref{17}, the induced line element becomes
\begin{equation}
	\begin{split}
	ds_{\xi_{st}} = \dfrac{\ell \left(\cos \left(\dfrac{l_{u}}{\ell}\right) - \cos \left(l_{\phi}\right)\right) r \: dr}{\sqrt{\left(r^2 \left(\cos \left(\dfrac{l_{u}}{\ell}\right) - \cos \left(l_{\phi}\right)\right) + \ell^{2} \cos \left(\dfrac{l_{u}}{\ell}\right)\right)^2 - \ell ^4}}
	\label{32}
\end{split}
\end{equation}
The length of the extremal curve is therefore
\begin{equation}
	\begin{split}
	L_{st} &= 2\int_{r_{\min}}^{\frac{\ell^2}{\varepsilon}} ds_{st}	=\ell \log \left[\dfrac{2 \ell^2}{\varepsilon^2} \left(\cos \left(\dfrac{l_{u}}{\ell}\right) -\cos \left(l_{\phi}\right)\right)\right] \\[0.4ex]{}&
	=\ell \log\left[\dfrac{4 \ell^2}{\varepsilon^2} \left(\sin \left(\dfrac{l_{\phi}}{2} + \dfrac{l_{u}}{2 \ell}\right) \sin \left(\dfrac{l_{\phi}}{2} - \dfrac{l_{u}}{2 \ell}\right)\right)\right].
	\label{53}
	\end{split}
\end{equation}

The entanglement entropy is obtained from the length of the extremal curve as
\begin{equation}
	\begin{aligned}
		S_{st} &= \dfrac{L_{st}}{4 G} 
		= \dfrac{\ell}{4 G} \log\left[\dfrac{4 \ell^2}{\varepsilon^2} \left(\sin \left(\dfrac{l_{\phi}}{2} + \dfrac{l_{u}}{2 \ell}\right) \sin \left(\dfrac{l_{\phi}}{2} - \dfrac{l_{u}}{2 \ell}\right)\right)\right] \\[0.4ex]
		&= \dfrac{c}{6} \log\left[\dfrac{4 \ell^2}{\varepsilon^2} \left(\sin \left(\dfrac{l_{\phi}}{2} + \dfrac{l_{u}}{2 \ell}\right) \sin \left(\dfrac{l_{\phi}}{2} - \dfrac{l_{u}}{2 \ell}\right)\right)\right] \\[0.4ex]
		&= \dfrac{c}{6} \log \biggl[\dfrac{\mathfrak{R}^2}{\pi^2 \varepsilon^2} \left(\sin \left(\dfrac{\pi}{\mathfrak{R}}\left(\Delta t + \Delta \phi\right)\right) \sin \left(\dfrac{\pi}{\mathfrak{R}} \left(\Delta t - \Delta \phi\right)\right)\right) \biggr].
	\end{aligned}
\end{equation}
Here, $\Delta t=\ell l_{\phi}$, $\Delta\phi=l_u$, and $\mathfrak{R}=2\pi\ell$. This result reproduces the real part of the timelike entanglement entropy for a timelike interval \cite{Doi_20231}.

\subsection{Imaginary Part}
\label{sec:imaginary_part}
Building on the spacelike branches constructed in the previous section, we now consider the timelike geodesic that completes the extremal curve associated with a timelike boundary interval. For the purely timelike interval defined in \eqref{9}, the corresponding boundary modular flow associated with the timelike geodesic reads
\begin{equation}
	\begin{split}
		\xi^{b}_{tt} = \dfrac{\pi}{T} \Bigl[ \left(L^{b}_{1} - \bar{L}^{b}_{1}\right) - T^2 \left(\bar{L}^{b}_{-1} - L^{b}_{-1}\right) \Bigr] 
		= -\dfrac{2 \pi t x}{T}\, \partial_{t} + \dfrac{\pi}{T}\left(-T^{2}-x^{2}-t^{2}\right) \partial_{x},
	\end{split}
\end{equation}
In addition to the modular flow $\xi^{b}_{tt}$, we can introduce another boundary vector field $\lambda^{b}_{tt}$ which is constructed from the same generators

\begin{equation}
\begin{split}
	\lambda^{b}_{tt} = \dfrac{\pi}{T} \Bigl[ \left(L^{b}_{1} + \bar{L}^{b}_{1}\right) + T^2 \left(L^{b}_{-1} + \bar{L}^{b}_{-1}\right) \Bigr] 
	= -\dfrac{2 \pi t x}{T}\, \partial_{x} + \dfrac{\pi}{T}\left(-T^{2}-x^{2}-t^{2}\right) \partial_{t},
\end{split}
\end{equation}
The bulk extensions of these flows are given by
\begin{equation}
	\begin{split}
		\xi^{B}_{tt} = \dfrac{\pi}{T} \Bigl[ \left(L^{B}_{1} - \bar{L}^{B}_{1}\right) - T^2 \left(\bar{L}^{B}_{-1} - L^{B}_{-1}\right) \Bigr] 
		= -\dfrac{2 \pi x}{T}\left(t \partial_{t} + z \partial_{z}\right) + \dfrac{\pi}{T}\left(-T^{2}-x^{2}-t^{2}+z^{2}\right) \partial_{x}
	\end{split}
	\label{46}
\end{equation}
\begin{equation}
\begin{split}
	\lambda^{B}_{tt} = \dfrac{\pi}{T} \Bigl[ \left(L^{B}_{1} + \bar{L}^{B}_{1}\right) + T^2 \left(L^{B}_{-1} + \bar{L}^{B}_{-1}\right) \Bigr] 
	= -\dfrac{2 \pi t}{T}\left(x \partial_{x} + z \partial_{z}\right) + \dfrac{\pi}{T}\left(-T^{2}-x^{2}-t^{2}-z^{2}\right) \partial_{t}.
\end{split}
\label{47}
\end{equation}

The modular flow $\xi^{B}_{tt}$ in \eqref{46} determines the extremal curve
\begin{equation}
	T^2 = z^2 - t^2, \qquad x = 0,
\end{equation}
which corresponds to the timelike extremal curve for a timelike interval and is in agreement with the known result in the literature \cite{Doi:2022iyj}; see Fig.~\ref{fig:1}.

\begin{figure}
	\centering
	\includegraphics[width=0.45\textwidth]{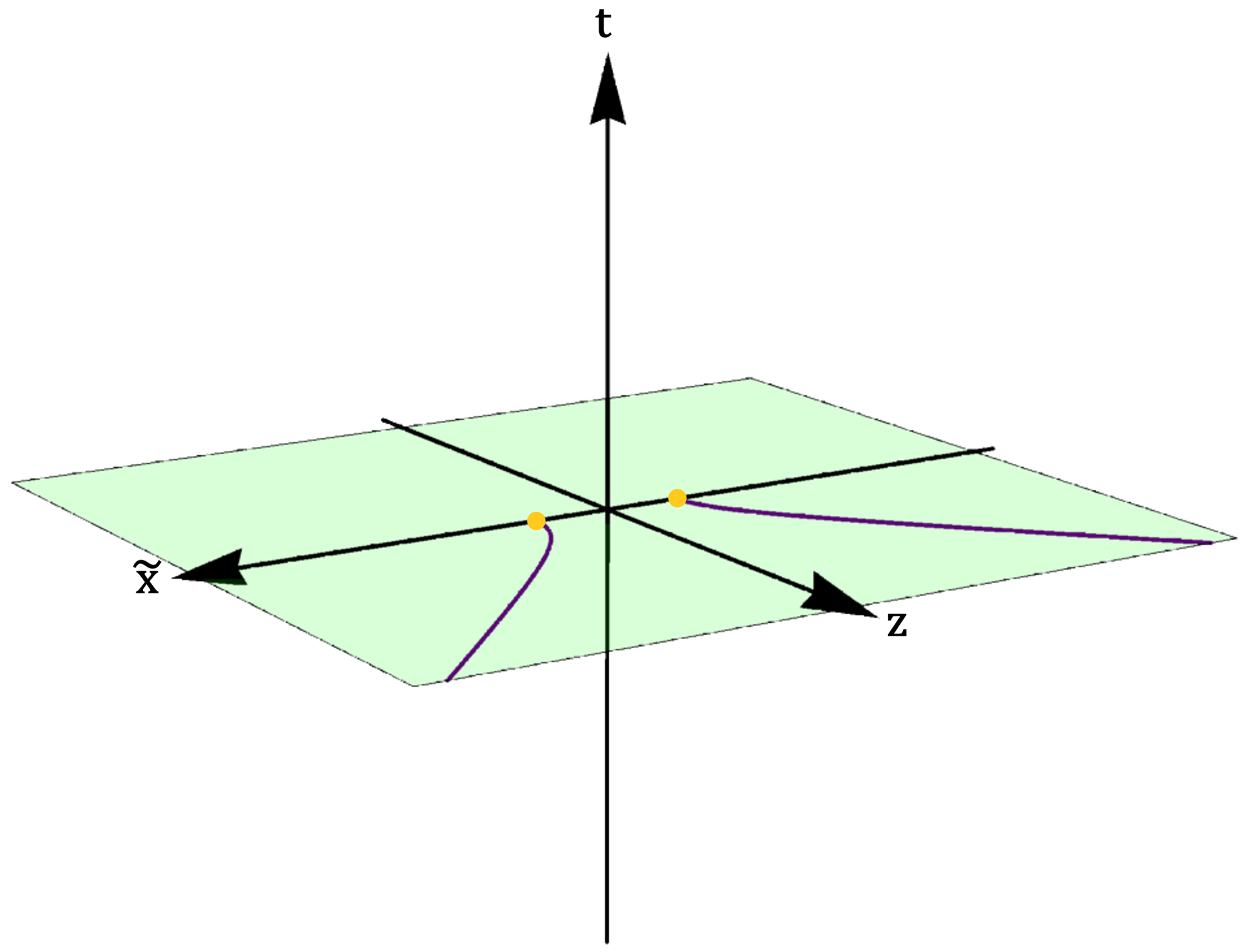}
	\caption{ \justifying Analytically continued curve with complexified embedding in the bulk. The two branches extend from the endpoints \((t, \tilde{x}, z)=(0, \pm T, 0)\) into the bulk.}
	\label{fig:2}
\end{figure}

For the vector field $\lambda^{B}_{tt}$ in \eqref{47}, the corresponding extremal curve is
\begin{equation}
	T^2 = - z^2 - x^2, \qquad t = 0.
	\label{60}
\end{equation}
It is worth noting that this curve does not admit a real solution in the bulk. However, upon analytic continuation of the embedding coordinate, we
obtain a novel result. We keep \(z\) real and positive, so that it continues to represent the bulk depth, we analytically continue the coordinate \(x\). Thus, we take
\begin{equation}
	x \rightarrow i\tilde{x}.
\end{equation}
With this continuation, the metric in \eqref{eq:poincare_metric} becomes
\begin{equation}
	d \tilde{s}^{2}=\dfrac{\ell^{2}}{z^{2}}(-d \tilde{x}^{2}+dz^{2}-dt^{2})
	\label{71}
\end{equation}
and the equations in \eqref{60} take the form
\begin{equation}
	T^2 = - z^2 + \tilde{x}^2, \qquad t = 0.
	\label{70}
\end{equation}
This relation describes a two branched curve with complexified embedding, extending into the bulk from the endpoints of the interval at $ - T<\tilde{x}<T$
as shown in Fig.~\ref{fig:2}.

Substituting \eqref{70} into the metric \eqref{71}, the corresponding line element becomes
\begin{equation}
	d \tilde{s} = \dfrac{\ell T dz}{z \sqrt{T^2 + z^2}}.
\end{equation}
Thus, we obtain
\begin{equation}
	\tilde{L} = 2 \int_{\epsilon}^{\infty} d \tilde{s} = 2 \ell \log \left(\dfrac{2 T}{\epsilon}\right).
\end{equation}
Although this curve is geometrically distinct from the two branched spacelike extremal curve associated with a timelike interval, its length precisely coincides with that of the latter \cite{Doi_20231}.

Extending the boundary construction to a generic interval through the boost transformation \eqref{7}, we obtain the corresponding boundary vector fields,
\begin{equation}
	\begin{split}
		\xi_{tt}^{b} = \dfrac{\pi \left(-l_{x}^{2} l_{t}+l_{t}^3 - 8 l_{x} x^{\prime} t^{\prime} + 4 l_{t}\left(x^{\prime2} + t^{\prime2}\right)\right)}{2 \left(l_{x}^2-l_{t}^2\right)} \partial_{t^{\prime}}  + \dfrac{\pi \left(l_{t}^{2} l_{x}-l_{x}^3 + 8 l_{t} x^{\prime} t^{\prime} - 4 l_{x}\left(x^{\prime2} + t^{\prime2}\right)\right)}{2 \left(l_{x}^2-l_{t}^2\right)} \partial_{x^{\prime}},
	\end{split}
\end{equation}
\begin{equation}
	\begin{split}
		\lambda_{tt}^{b} = \dfrac{\pi \left(-l_{t}^{2} l_{x}+l_{x}^3 - 8 l_{t} x^{\prime} t^{\prime} + 4 l_{x}\left(x^{\prime2} + t^{\prime2}\right)\right)}{2 \left(l_{t}^2-l_{x}^2\right)} \partial_{t^{\prime}} + \dfrac{\pi \left(l_{x}^{2} l_{t}-l_{t}^3 + 8 l_{x} x^{\prime} t^{\prime} - 4 l_{t}\left(x^{\prime2} + t^{\prime2}\right)\right)}{2 \left(l_{t}^2-l_{x}^2\right)} \partial_{x^{\prime}}.
	\end{split}
\end{equation}
The associated bulk vector fields are then obtained as
\begin{equation}
	\begin{split}
		\xi_{tt}^{B} &= \dfrac{\pi \left(-l_{x}^{2} l_{t}+l_{t}^3 - 8 l_{x} x^{\prime} t^{\prime} + 4 l_{t}\left(x^{\prime2} + t^{\prime2} + z^{2}\right)\right)}{2 \left(l_{x}^2-l_{t}^2\right)} \partial_{t^{\prime}} \\{}& \quad + \dfrac{\pi \left(l_{t}^{2} l_{x}-l_{x}^3 + 8 l_{t} x^{\prime} t^{\prime} - 4 l_{x}\left(x^{\prime2} + t^{\prime2} - z^{2}\right)\right)}{2 \left(l_{x}^2-l_{t}^2\right)} \partial_{x^{\prime}} \\{}& \quad - \dfrac{4 \pi \left(l_{x} x^{\prime} - l_{t} t^{\prime}\right) z}{l_{x}^{2}-l_{t}^2} \partial_{z}
	\end{split}
	\label{14}
\end{equation}

\begin{equation}
	\begin{split}
		\lambda_{tt}^{B} &= \dfrac{\pi \left(-l_{t}^{2} l_{x}+l_{x}^3 - 8 l_{t} x^{\prime} t^{\prime} + 4 l_{x}\left(x^{\prime2} + t^{\prime2} + z^{2}\right)\right)}{2 \left(l_{t}^2-l_{x}^2\right)} \partial_{t^{\prime}} \\{}& \quad + \dfrac{\pi \left(l_{x}^{2} l_{t}-l_{t}^3 + 8 l_{x} x^{\prime} t^{\prime} - 4 l_{t}\left(x^{\prime2} + t^{\prime2} - z^{2}\right)\right)}{2 \left(l_{t}^2-l_{x}^2\right)} \partial_{x^{\prime}} \\{}& \quad - \dfrac{4 \pi \left(l_{t} x^{\prime} - l_{x} t^{\prime}\right) z}{l_{t}^{2}-l_{x}^2} \partial_{z}
	\end{split}
	\label{15}
\end{equation}
The above bulk flows reveal a simple correspondence: the transformation $l_{x}\leftrightarrow l_{t}$ exchanges the two vector fields.
Furthermore, these flows obey the following  relations
\begin{equation}
	\begin{split}
	\left(\xi_{tt}^{B}\right)^{\mu} \left(\lambda_{tt}^{B}\right)_{\mu} &= 0 \\
	\left(\xi_{tt}^{B}\right)^{\mu} \left(\xi_{tt}^{B}\right)_{\mu}  + \left(\lambda_{tt}^{B}\right)^{\mu} \left(\lambda_{tt}^{B}\right)_{\mu} &= - 4 \pi^{2} \ell^2,
	\end{split}
\end{equation}
which means that $\xi_{tt}^{B}$ and $\lambda_{tt}^{B}$ are orthogonal to each other and satisfy a novel norm relation that is like its spacelike counterpart in \eqref{4} except for a minus sign.

The extremal curve associated with the modular flow  $\xi_{tt}^{B}$ in \eqref{14} is given by
\begin{equation}
	x^{\prime2} = \dfrac{l_{t}^{2}(l_{t}^2-l_{x}^{2} + 4z^{2})}{4(l_{x}^{2}-l_{t}^{2})} , \qquad t^{\prime2} = \dfrac{l_{x}^{2}(l_{t}^2-l_{x}^{2}+4z^{2})}{4(l_{x}^{2}-l_{t}^{2})} .
\end{equation}
The resulting curve coincides with the known timelike extremal curve \cite{Fareghbal_2026}.
For the vector field  $\lambda_{tt}^{B}$ in \eqref{15}, the corresponding extremal curve reads
\begin{equation}
	x^{\prime2} = \dfrac{l_{x}^{2}(l_{x}^2-l_{t}^{2} + 4z^{2})}{4(l_{t}^{2}-l_{x}^{2})} , \qquad t^{\prime2} = \dfrac{l_{t}^{2}(l_{x}^2-l_{t}^{2}+4z^{2})}{4(l_{t}^{2}-l_{x}^{2})} .
\end{equation}

To express the resulting boundary vector fields in BMS coordinates, we again employ the BMS transformation \eqref{8} together with the relations between the interval parameters given in \eqref{11}, which yields
\begin{widetext}
		\begin{equation}
			\begin{split}
				\xi_{tt}^{b} &= \dfrac{2 \pi \ell \left(- 2 \cos \left(\dfrac{l_{\phi}}{2}\right) \sin \left(\dfrac{l_{u}}{2 \ell}\right) + \cos \left(\dfrac{u}{\ell}\right) \cos \left(\phi\right) \sin \left(\dfrac{l_{u}}{\ell}\right) + \sin \left(l_{\phi}\right) \sin \left(\dfrac{u}{\ell}\right) \sin \left(\phi\right)\right)}{\cos \left(l_{\phi}\right) - \cos \left(\dfrac{l_{u}}{\ell}\right)} \partial_{u} \\{}& \quad - \dfrac{2 \pi \left(-2 \cos \left(\dfrac{l_{u}}{2 \ell}\right) \sin \left(\dfrac{l_{\phi}}{2}\right) + \cos \left(\phi\right) \sin \left(l_{\phi}\right) \cos \left(\dfrac{u}{\ell}\right) + \sin \left(\dfrac{l_{u}}{\ell}\right) \sin \left(\phi\right) \sin \left(\dfrac{u}{\ell}\right)\right)}{\cos \left(l_{\phi}\right) - \cos \left(\dfrac{l_{u}}{\ell}\right)} \partial_{\phi},
			\end{split}
			\label{50}
		\end{equation}
		\begin{equation}
			\begin{split}
				\lambda_{tt}^{b} &= - \dfrac{2 \pi \ell \left(- 2 \cos \left(\dfrac{l_{u}}{2 \ell}\right) \sin \left(\dfrac{l_{\phi}}{2}\right) + \cos \left(\dfrac{u}{\ell}\right) \cos \left(\phi\right) \sin \left(l_{\phi}\right) + \sin \left(\dfrac{l_{u}}{\ell}\right) \sin \left(\dfrac{u}{\ell}\right) \sin \left(\phi\right)\right)}{\cos \left(l_{\phi}\right) - \cos \left(\dfrac{l_{u}}{\ell}\right)} \partial_{u} \\{}& \quad + \dfrac{2 \pi \left(-2 \cos \left(\dfrac{l_{\phi}}{2}\right) \sin \left(\dfrac{l_{u}}{2 \ell}\right) + \cos \left(\phi\right) \sin \left(\dfrac{l_{u}}{\ell}\right) \cos \left(\dfrac{u}{\ell}\right) + \sin \left(l_{\phi}\right) \sin \left(\phi\right) \sin \left(\dfrac{u}{\ell}\right)\right)}{\cos \left(l_{\phi}\right) - \cos \left(\dfrac{l_{u}}{\ell}\right)} \partial_{\phi}.
			\end{split}
		\end{equation}
\end{widetext}
The bulk counterparts of these boundary flows are~given~by
\begin{widetext}
	\begin{scriptsize}
		\begin{equation}
			\begin{split}
				\xi_{tt}^{B} &= \dfrac{2 \pi \ell \left(- 2 \cos \left(\dfrac{l_{\phi}}{2}\right) \sin \left(\dfrac{l_{u}}{2 \ell}\right) + \cos \left(\dfrac{u}{\ell}\right) \cos \left(\phi\right) \sin \left(\dfrac{l_{u}}{\ell}\right) + \sin \left(l_{\phi}\right) \sin \left(\dfrac{u}{\ell}\right) \sin \left(\phi\right)\right)}{\cos \left(l_{\phi}\right) - \cos \left(\dfrac{l_{u}}{\ell}\right)} \partial_{u} \\[0.7ex]{}& \quad - \dfrac{2 \pi \left(\cos \left(\phi\right) \sin \left(\dfrac{l_{u}}{\ell}\right) \left(\ell \cos \left(\dfrac{u}{\ell}\right) - r \sin \left(\dfrac{u}{\ell}\right)\right) + \sin \left(l_{\phi}\right) \sin \left(\phi\right) \left(r \cos \left(\dfrac{u}{\ell}\right) + \ell \sin \left(\dfrac{u}{\ell}\right)\right)\right)}{\cos \left(l_{\phi}\right) - \cos \left(\dfrac{l_{u}}{\ell}\right)} \partial_{r} \\[0.7ex]{}& \quad - \dfrac{2 \pi \left(-2 r \cos \left(\dfrac{l_{u}}{2 \ell}\right) \sin \left(\dfrac{l_{\phi}}{2}\right) + \cos \left(\phi\right) \sin \left(l_{\phi}\right) \left(\ell \sin \left(\dfrac{u}{\ell}\right) + r \cos \left(\dfrac{u}{\ell}\right)\right) + \sin \left(\dfrac{l_{u}}{\ell}\right) \sin \left(\phi\right) \left(r \sin \left(\dfrac{u}{\ell}\right) - \ell \cos \left(\dfrac{u}{\ell}\right)\right)\right)}{r \left(\cos \left(l_{\phi}\right) - \cos \left(\dfrac{l_{u}}{\ell}\right)\right)} \partial_{\phi},
			\end{split}
			\label{48}
		\end{equation}
	\end{scriptsize}

	\begin{scriptsize}
		\begin{equation}
			\begin{split}
				\lambda_{tt}^{B} &= - \dfrac{2 \pi \ell \left(- 2 \cos \left(\dfrac{l_{u}}{2 \ell}\right) \sin \left(\dfrac{l_{\phi}}{2}\right) + \cos \left(\dfrac{u}{\ell}\right) \cos \left(\phi\right) \sin \left(l_{\phi}\right) + \sin \left(\dfrac{l_{u}}{\ell}\right) \sin \left(\dfrac{u}{\ell}\right) \sin \left(\phi\right)\right)}{\cos \left(l_{\phi}\right) - \cos \left(\dfrac{l_{u}}{\ell}\right)} \partial_{u} \\[0.7ex]{}& \quad + \dfrac{2 \pi \left(\cos \left(\phi\right) \sin \left(l_{\phi}\right) \left(\ell \cos \left(\dfrac{u}{\ell}\right) - r \sin \left(\dfrac{u}{\ell}\right)\right) + \sin \left(\dfrac{l_{u}}{\ell}\right) \sin \left(\phi\right) \left(r \cos \left(\dfrac{u}{\ell}\right) + \ell \sin \left(\dfrac{u}{\ell}\right)\right)\right)}{\cos \left(l_{\phi}\right) - \cos \left(\dfrac{l_{u}}{\ell}\right)} \partial_{r} \\[0.7ex]{}& \quad + \dfrac{2 \pi \left(-2 r \cos \left(\dfrac{l_{\phi}}{2}\right) \sin \left(\dfrac{l_{u}}{2 \ell}\right) + \cos \left(\phi\right) \sin \left(\dfrac{l_{u}}{\ell}\right) \left(\ell \sin \left(\dfrac{u}{\ell}\right) + r \cos \left(\dfrac{u}{\ell}\right)\right) + \sin \left(l_{\phi}\right) \sin \left(\phi\right) \left(r \sin \left(\dfrac{u}{\ell}\right) - \ell \cos \left(\dfrac{u}{\ell}\right)\right)\right)}{r \left(\cos \left(l_{\phi}\right) - \cos \left(\dfrac{l_{u}}{\ell}\right)\right)} \partial_{\phi}.
			\end{split}
			\label{49}
		\end{equation}
	\end{scriptsize}
\end{widetext}
As before,  the symmetry relations $l_{u} \rightarrow \ell l_{\phi}$ and $ l_{\phi} \rightarrow l_{u}/ \ell$ interchange $\xi_{tt}^{b}$ and $\lambda_{tt}^{b}$.

The modular flow $\xi^{B}_{tt}$ in \eqref{48} gives rise to the extremal curve 
\begin{equation}
	\begin{split}
	r = \dfrac{\ell \cos \left(\dfrac{l_{\phi}}{2}\right) \sin \left(\dfrac{u}{\ell}\right)}{\cos \left(\dfrac{l_{u}}{2 \ell}\right) \cos \left(\phi\right) - \cos \left(\dfrac{l_{\phi}}{2}\right) \cos \left(\dfrac{u}{\ell}\right)}, \\
	-r \sin \left(\phi\right) \sin \left(\dfrac{l_{\phi}}{2}\right) = \sin \left(\dfrac{l_{u}}{2 \ell}\right) (\ell \cos \left(\dfrac{u}{\ell}\right) - r \sin \left(\dfrac{u}{\ell}\right)),
	\end{split}
	\label{18}
\end{equation}
in accordance with the extremal curve obtained in \cite{Fareghbal_2026}. On the other hand, the vector field $\lambda^{B}_{tt}$ in \eqref{49} leads to the extremal curve 

\begin{equation}
		\begin{split}
	r = \dfrac{\ell \cos \left(\dfrac{l_{u}}{2 \ell}\right) \sin \left(\dfrac{u}{\ell}\right)}{\cos \left(\dfrac{l_{\phi}}{2}\right) \cos \left(\phi\right) - \cos \left(\dfrac{l_{u}}{2 \ell}\right) \cos \left(\dfrac{u}{\ell}\right)}, \\
	-r \sin \left(\phi\right) \sin \left(\dfrac{l_{u}}{2 \ell}\right) = \sin \left(\dfrac{l_{\phi}}{2}\right) (\ell \cos \left(\dfrac{u}{\ell}\right) - r \sin \left(\dfrac{u}{\ell}\right)).
\end{split}
\end{equation}
For later convenience, we collect the two relations in \eqref{18} to write $r^{2}$ as follows:
\begin{equation}
	\begin{split}
		r^{2} = \dfrac{4 \ell^2 \cos^{2} \left(\dfrac{l_{\phi}}{2}\right) \sin^{2} \left(\dfrac{l_{u}}{2 \ell}\right)}{\left(\cos \left(\dfrac{l_{u}}{\ell}\right) - \cos \left(l_{\phi}\right)\right) \left(1 - \cos \left(\dfrac{l_{u}}{\ell}\right) \cos^{2} \left(\phi\right) + \cos \left(l_{\phi}\right) \sin^{2} \left(\phi\right)\right)}.
	\end{split}
\end{equation}

Since this timelike curve connects the two branches of the spacelike extremal curve, its minimum radial coordinate coincides with the joining point of the spacelike branches, which is given by $r_{\min}$ as in \eqref{31}. The maximum radial coordinate, on the other hand, is obtained by extremizing the above relation, yielding
\begin{equation}
	r_{\max} =  \dfrac{\ell \cos \left(\dfrac{l_{\phi}}{2}\right)}{\sqrt{\sin^{2} \left(\dfrac{l_{\phi}}{2}\right) - \sin^{2}\left(\dfrac{l_{u}}{2 \ell}\right)}}.
\end{equation}

Combining the extremal curve equations in \eqref{18} with the bulk metric in \eqref{17}, the corresponding line element is
\begin{widetext}
\begin{footnotesize} 
	\begin{equation}
		ds_{\xi_{tt}} = \dfrac{ i \ell \left(\cos \left(\dfrac{l_{u}}{\ell}\right) - \cos \left(l_{\phi}\right)\right) r \: dr}{\sqrt{\left(\ell ^2 \left(1 + \cos \left(l_{\phi}\right)\right) - r^2 \left(\cos \left(\dfrac{l_{u}}{\ell}\right) - \cos \left(l_{\phi}\right)\right)\right)  \left(r^2 \left(\cos \left(\dfrac{l_{u}}{\ell}\right) - \cos \left(l_{\phi}\right)\right) - \ell^2 \left(1 - \cos \left(\dfrac{l_{u}}{\ell}\right)\right)\right)}}
		\label{33}
	\end{equation}
\end{footnotesize}
\end{widetext} 
Evaluating the length of this curve gives
\begin{equation}
	L_{tt} = 2 \int_{r_{\min}}^{r_{\max}} ds_{\xi_{tt}} = i  \pi \ell.
	\label{54}
\end{equation}
In the Rindler picture of \cite{shalamzari2026holographicpseudoentropyrindler}, the flow $\xi_{tt}$
vanishes on a timelike geodesic stretched between the inner and outer
horizons of the BTZ black hole, and its length $i\pi\ell$ is
independent of the interval and of the horizon radii.
The imaginary part of the holographic timelike entanglement entropy is therefore
\begin{equation}
	S_{tt} = \dfrac{L_{tt}}{4 G} = \dfrac{i \pi \ell}{4 G} = \dfrac{i \pi c}{6}.
\end{equation}
Upon summing the real and imaginary contributions, the holographic entanglement entropy for a timelike interval takes the form \cite{Doi_20231} 
\begin{equation}
	\begin{split}
		S^{(T)} &= \dfrac{L_{st}+L_{tt}}{4 G} 
		= \dfrac{c}{6} \log\left[\dfrac{4 \ell^2}{\varepsilon^2} \left(\sin \left(\dfrac{l_{\phi}}{2} + \dfrac{l_{u}}{2 \ell}\right) \sin \left(\dfrac{l_{\phi}}{2} - \dfrac{l_{u}}{2 \ell}\right)\right)\right]  +\dfrac{i \pi c}{6}.
	\end{split}
\end{equation}
The real part reproduces the standard logarithmic form of the entanglement entropy on the cylinder. The imaginary part is a constant, $i\pi c/6$, fixed entirely by the central charge and independent of the interval size, consistent with previous results in the literature \cite{Doi:2022iyj,Doi_20231}.

\section{Timelike Entanglement Entropy in Flat Space from Modular Flow}
\label{Timelike Entanglement Entropy in Flat Space from Modular Flow}

We turn next to flat space holography, extending the AdS$_3$/CFT$_2$ construction of the previous two sections. The relevant boundary modular flows are constructed from the CCFT$_2$ symmetry generators. The corresponding bulk symmetry generators are Killing vectors of flat spacetime. These generators yield bulk modular flows for both spacelike and timelike intervals, whose fixed points determine the corresponding extremal curves and, in turn, the holographic (timelike) entanglement entropy. As in AdS$_3$, the timelike entanglement entropy again splits into a real and an imaginary part, associated respectively with spacelike and timelike extremal curves. We then compare these flat space results with the flat limit of the corresponding AdS quantities. We note that although the modular flows $\xi_{ss}^{b}$ and $\xi_{ss}^{B}$, associated with a spacelike boundary interval, do not admit a well-defined flat limit, the alternative flows $\xi_{st}^b$ and $\xi_{st}^B$ do. The flat limits of $\xi_{st}^b$ and $\xi_{st}^B$ are identical to the flows $\zeta_{st}^b$ and $\zeta_{st}^B$, respectively, constructed below for the timelike interval. Consequently, the extremal curves, the induced line element, and the resulting length are identical to those obtained below; the only difference lies in the choice of the boundary interval. We therefore refrain from repeating this calculation for the spacelike interval.

\subsection{Real Part}
We consider flat spacetime in global Minkowski coordinates, in which the metric is given by 
\begin{equation}
	ds^{2} = -du^{2} -2 du dr + r^{2} d\phi^{2}.
	\label{12}
\end{equation}
where $u$ and $r$ denote the retarded time and radial coordinate, respectively. In these coordinates, the globally well-defined boundary CCFT$_{2}$ symmetry generators on the cylinder take the form: 
\begin{equation}
	\begin{aligned}
		L^{b}_{-1} &= e^{i\phi}\left(u\partial_u-i\partial_\phi\right),
		&\hspace{1cm} M^{b}_{-1} &= i e^{i\phi}\partial_u, \\[0.3cm]
		L^{b}_{0} &= -i\partial_\phi,
		&\hspace{1cm} M^{b}_{0} &= i\partial_u, \\[0.3cm]
		L^{b}_{1} &= -e^{-i\phi}\left(u\partial_u+i\partial_\phi\right),
		&\qquad M^{b}_{1} &= i e^{-i\phi}\partial_u.
	\end{aligned}
		\label{75}
\end{equation}
The bulk extensions of these generators are isometry group generators of three-dimensional flat space time that are given by
\begin{equation}
	\begin{split}
		L^{B}_{-1} &= e^{i\phi} \left(u\partial_u-(u+r)\partial_r - i\left(1+\dfrac{u}{r}\right)\partial_\phi\right), \\[0.4ex]
		L^{B}_{0} &= -i\partial_\phi, \\[0.4ex]
		L^{B}_{1} &= -e^{-i\phi} \left(u\partial_u-(u+r)\partial_r + i\left(1+\dfrac{u}{r}\right)\partial_\phi\right), \\[1ex]
		M^{B}_{-1} &= e^{i\phi} \left(i\partial_u-i\partial_r+\dfrac{1}{r}\partial_\phi\right), \\[0.4ex]
		M^{B}_{0} &= i\partial_u, \\[0.4ex]
		M^{B}_{1} &= e^{-i\phi} \left(i\partial_u-i\partial_r-\dfrac{1}{r}\partial_\phi\right).
	\end{split}
	\label{76}
\end{equation}
We consider a timelike interval on the null boundary with parameters $l_{u}$ and $l_{\phi}$. It is called timelike in the sense that the modular flows constructed below will be shown to coincide with the flat limit of the modular flows of the timelike interval in AdS$_3$.

With the symmetry generators, \eqref{75}, the associated boundary modular flow reads
\begin{widetext}
		\begin{equation}
			\begin{split}
				\zeta^{b}_{st} &= i \pi \csc \dfrac{l_{\phi}}{2} \left( \left(L^{b}_{1} + L^{b}_{-1}\right) + \dfrac{l_{u}}{2} \cot \dfrac{l_{\phi}}{2} \left(M^{b}_{1} + M^{b}_{-1}\right) - l_{u} \csc \dfrac{l_{\phi}}{2} M^{b}_{0} - 2 \cos \dfrac{l_{\phi}}{2} L^{b}_{0}\right) \\[0.7ex]
				{}&= - \pi \csc \left( \dfrac{l_{\phi}}{2} \right) \left(l_{u} \cos \left(\phi \right) \cot \left(\dfrac{l_{\phi}}{2} \right) + 2 u \sin \left( \phi \right) - l_{u} \csc \left(\dfrac{l_{\phi}}{2} \right)\right) \partial_{u} \\[0.7ex]
				{}& \quad - 2 \pi \csc \left(\dfrac{l_{\phi}}{2}\right) \left(\cos \left(\dfrac{l_{\phi}}{2}\right) - \cos \left(\phi\right)\right) \partial_{\phi}.
			\end{split}
		\end{equation}
\end{widetext}
The boundary flow vanishes at the two endpoints of the interval, $\left(u,\phi\right)=\left(l_u/2,\,l_\phi/2\right)$ and $\left(-l_u/2,\,-l_\phi/2\right)$.
Using the bulk generators, \eqref{76}, the modular flow admits a bulk extension of the form
\begin{widetext}
		\begin{equation}
			\begin{split}
				\zeta^{B}_{st} &= i \pi \csc \dfrac{l_{\phi}}{2} \biggl( \left(L^{B}_{1} + L^{B}_{-1}\right) + \dfrac{l_{u}}{2} \cot \dfrac{l_{\phi}}{2} \left(M^{B}_{1} + M^{B}_{-1}\right) - l_{u} \csc \dfrac{l_{\phi}}{2} M^{B}_{0} - 2 \cos \dfrac{l_{\phi}}{2} L^{B}_{0} \biggr) \\[0.7ex]
				{}&= - \pi \csc \left( \dfrac{l_{\phi}}{2} \right) \left(l_{u} \cos \left(\phi \right) \cot \left(\dfrac{l_{\phi}}{2} \right) + 2 u \sin \left( \phi \right) - l_{u} \csc \left(\dfrac{l_{\phi}}{2} \right)\right) \partial_{u} \\[0.7ex]
				{}& \quad + \pi \csc \left(\dfrac{l_{\phi}}{2}\right) \left(l_{u} \cos \left(\phi\right) \cot \left(\dfrac{l_{\phi}}{2}\right) + 2(r + u) \sin \left(\phi\right) \right) \partial_{r} \\[0.7ex]
				{}& \quad - \pi \csc \left(\dfrac{l_{\phi}}{2}\right) \left( 2\left(\cos \left(\dfrac{l_{\phi}}{2}\right) - \cos \left(\phi\right)\right) + \dfrac{1}{r} \left(l_{u} \sin \left(\phi\right) \cot \left(\dfrac{l_{\phi}}{2}\right) - 2 u \cos \left(\phi\right)\right)\right) \partial_{\phi}.
			\end{split}
			\label{21}
		\end{equation}
\end{widetext}
It is important to note that the modular flows derived in flat spacetime can also be recovered from the AdS$_3$ results in the flat space limit. In particular, by taking the limit $\ell \rightarrow\infty$ in \eqref{77} and \eqref{44}, the boundary and bulk modular flows reduce precisely to the expressions obtained in   the context of flat space holography.

Starting from the bulk modular flow $\zeta^{B}_{st}$, the flat spacelike extremal curve is obtained as follows
\begin{equation}
	\begin{split}
	r \sin \left(\phi\right) \sin \left(\dfrac{l_{\phi}}{2}\right) = - \dfrac{l_{u}}{2}, \\
	r+ u = r \cos \left(\phi\right) \cos \left(\dfrac{l_{\phi}}{2}\right).
	\end{split}
	\label{13}
\end{equation}
The flat limit of the AdS$_3$  spacelike extremal curve in \eqref{16} reproduces this spacelike extremal curve \cite{Fareghbal_2026}. 
A convenient choice of coordinates allows the modular flow to take a simpler form. The transformations \cite{Fareghbal:2019czx} are given by 
\begin{equation}
	\begin{split}
		t &= \left(r + u\right) \csc \left(\dfrac{l_{\phi}}{2}\right) - r \cos \left(\phi\right) \cot \left(\dfrac{l_{\phi}}{2}\right), \\[0.7ex]
		x &= r \sin \left(\phi\right) + \dfrac{l_{u}}{2} \csc \left(\dfrac{l_{\phi}}{2}\right), \\[0.7ex]
		y &= r \cos \left(\phi\right) \csc \left(\dfrac{l_{\phi}}{2}\right) - \left(r + u\right) \cot \left(\dfrac{l_{\phi}}{2}\right).
	\end{split}
	\label{56}
\end{equation}
The inverse transformation can be written as
\begin{footnotesize}
	\begin{equation}
		r = \sqrt{\left(x - \dfrac{l_{u}}{2}\csc \left(\dfrac{l_{\phi}}{2}\right) \right)^2 + \left(t \cot \left(\dfrac{l_{\phi}}{2}\right) +y \csc \left(\dfrac{l_{\phi}}{2}\right)\right)^2},
		\label{57}
	\end{equation}
	\begin{equation*}
		\phi = \sin^{-1}\left( \dfrac{x - \dfrac{l_{u}}{2} \csc \left(\dfrac{l_{\phi}}{2}\right)}{\sqrt{\left(x - \dfrac{l_{u}}{2}\csc \left(\dfrac{l_{\phi}}{2}\right) \right)^2 + \left(t \cot \left(\dfrac{l_{\phi}}{2}\right) +y \csc \left(\dfrac{l_{\phi}}{2}\right)\right)^2}}\right),
	\end{equation*}
\begin{equation*}
	\begin{split}
		u &= \dfrac{1}{2} \left( \cot \left(\dfrac{l_{\phi}}{4}\right) \left(t + y\right) + \tan \left(\dfrac{l_{\phi}}{4}\right) \left(t - y\right) \right) - \sqrt{\left(x - \dfrac{l_{u}}{2}\csc \left(\dfrac{l_{\phi}}{2}\right) \right)^2 + \left(t \cot \left(\dfrac{l_{\phi}}{2}\right) +y \csc \left(\dfrac{l_{\phi}}{2}\right)\right)^2}.
	\end{split}
\end{equation*}
\end{footnotesize}
In these coordinates, the metric \eqref{12} takes the Cartesian form
\begin{equation}
	ds^{2} = - dt^{2} + dx^{2} + dy^{2}
\end{equation}
Thus, the bulk modular flow becomes
\begin{equation}
	\zeta^{B}_{st} = 2 \pi \left(x \partial_{t} + t \partial_{x}\right).
	\label{25}
\end{equation}

	In these Cartesian coordinates the geometry of the two flows is simple.
The flow $\zeta^{B}_{st}$ is a boost in the $(t,x)$ plane. Its norm,
$4\pi^2(t^2-x^2)$, vanishes on the null planes $x=\pm t$, which form a
Rindler horizon of flat space, and its fixed points $x=t=0$ form the
bifurcation line. Using \eqref{13} and \eqref{56}, the spacelike
extremal curve is exactly this line. In the same way, the flow
$\zeta^{B}_{tt}=2\pi(x\partial_y-y\partial_x)$ is a rotation in the
$(x,y)$ plane, with norm $4\pi^2(x^2+y^2)$, and its fixed points
$x=y=0$ form a timelike line along $t$. Using \eqref{55} and
\eqref{56}, the timelike extremal curve is this line. The two curves
are orthogonal and meet at the origin. The relation of this picture to
the horizons of the flat-space cosmology, the flat limit of the inner
and outer horizons of BTZ, and the case of the plane are discussed in
the companion paper \cite{shalamzari2026holographicpseudoentropyrindler}. 
In flat space holography, the entire extremal curve does not contribute to the entanglement entropy. Instead, only the segment of the curve lying between the two null rays contributes. These null rays emanate from the two endpoints of the boundary interval at null infinity and extend into the bulk, where they intersect the spacelike extremal curve at two distinct points \cite{Jiang_2017}. The locations of these intersection points can be determined as follows.

\begin{figure}
	\centering
	\includegraphics[width=0.55\textwidth]{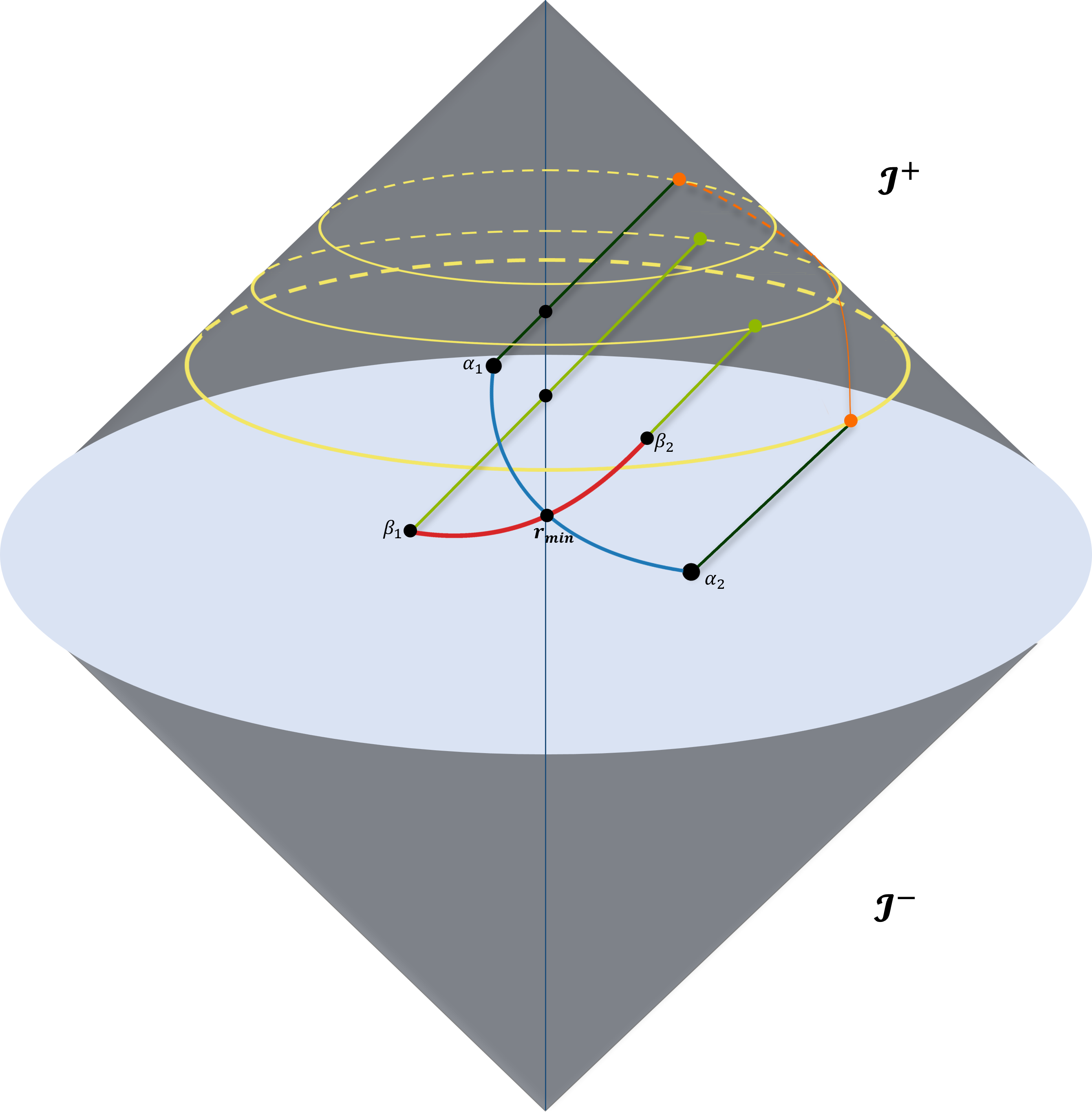}
	\caption{ \justifying The spacelike and timelike extremal curves and their corresponding null rays. The boundary interval at null infinity is shown in orange, while the spacelike and timelike extremal curves are shown in blue and red, respectively. The null rays corresponding to the spacelike and timelike extremal curves are shown in dark green and light green, respectively.}
	
	\label{fig:3}
\end{figure}

From the first equation in \eqref{13}, the minimum value of the radial coordinate is immediately obtained as
\begin{equation}
	r_{\min}^{f} = \dfrac{l_{u}}{2} \csc\left(\dfrac{l_{\phi}}{2}\right).
	\label{27}
\end{equation}
By combining the two equations in \eqref{13}, we derive the following equation for the extremal curve

\begin{equation}
	u = \sqrt{r^{2} \cos^{2} \dfrac{l_{\phi}}{2}
		- \dfrac{l_{u}^{2}}{4} \cot^{2} \dfrac{l_{\phi}}{2}} - r.
\end{equation}
The retarded time of the second intersection point has an extremum value at the radius $r_{I}^{f}$. Thus, imposing $\dfrac{du}{dr}=0$, we find
\begin{equation}
	\label{79}
	r_{I}^{f} = \dfrac{l_{u}}{2} \csc^{2}\left(\dfrac{l_{\phi}}{2}\right).
\end{equation}
Substituting \(r=r_{I}^{f}\) into the spacelike extremal curve equations, the two intersection points are found to be
\begin{equation}
	\begin{aligned}
		\alpha_{1} =	(u_1,\phi_1,r_I^{f}) &= ( \dfrac{l_{u}}{2} - l_u \csc^2 \dfrac{l_\phi}{2}, \dfrac{l_\phi}{2} - \pi ,r_I^{f}),\\
		\alpha_{2} =	(u_2,\phi_2,r_I^{f}) &= (-\dfrac{l_u}{2},-\dfrac{l_\phi}{2},r_I^{f}).
	\end{aligned}
\end{equation}
The spacelike extremal curve and the two null rays connecting it to the endpoints of the boundary interval are shown in Fig.~\ref{fig:3}.

Inserting the curve equations \eqref{13} into the metric \eqref{12} gives the induced line element
\begin{equation}
	ds^{f}_{\zeta_{st}} = \dfrac{2\: r \sin^{2} \left(\dfrac{l_{\phi}}{2}\right) }{\sqrt{2 r^2 \left(1 - \cos \left(l_{\phi}\right)\right) - l_{u}^2}} \: dr
	\label{28}
\end{equation}
Thus,
\begin{equation}
	L_{st}^{f} = 2 \int_{r_{\min}^{}}^{r_{I}^{f}} ds^{f}_{\zeta_{st}} = l_{u} \cot \left(\dfrac{l_{\phi}}{2}\right).
	\label{29}
\end{equation}
This length divided by $4G$ gives rise to the real part of the timelike entanglement entropy. The real part reproduces the entanglement entropy that has been computed before in CCFT$_{2}$~\cite{Jiang_2017}.

\subsection{Imaginary Part}
Using the generators introduced in \eqref{75}, we construct the boundary modular flow, which is given by
\begin{widetext}
		\begin{equation}
			\begin{split}
				\zeta^{b}_{tt} &= i \pi \csc \left(\dfrac{l_{\phi}}{2}\right) \biggl( \cos \left(\dfrac{l_{\phi}}{2}\right) \left(L^{b}_{1} + L^{b}_{-1}\right) + \dfrac{l_{u}}{2} \csc \left(\dfrac{l_{\phi}}{2}\right) \left(M^{b}_{1} + M^{b}_{-1}\right) - l_{u} \cot \left(\dfrac{l_{\phi}}{2}\right) M^{b}_{0} - 2 L^{b}_{0} \biggr) \\[0.7ex]
				{}&=- \pi \csc \left(\dfrac{l_{\phi}}{2}\right) \left(l_{u} \cos \left(\phi\right) \csc \left(\dfrac{l_{\phi}}{2}\right) + 2 u \sin \left(\phi\right) \cos \left(\dfrac{l_{\phi}}{2}\right) - l_{u} \cot \left(\dfrac{l_{\phi}}{2}\right)\right) \partial_{u} \\[0.7ex]
				{}& \quad + \pi \csc \left(\dfrac{l_{\phi}}{2}\right) \left(2 \cos \left(\dfrac{l_{\phi}}{2}\right) \cos \left(\phi\right) - 2\right) \partial_{\phi}.
			\end{split}
			\label{78}
		\end{equation}
\end{widetext}
Unlike the boundary modular flow $\zeta_{st}^{b} $, the above modular flow does not vanish at any point on the boundary. The bulk extension of this modular flow can be written as
\begin{widetext}
		\begin{equation}
			\begin{split}
				\zeta^{B}_{tt} &= i \pi \csc \left(\dfrac{l_{\phi}}{2}\right) \biggl( \cos \left(\dfrac{l_{\phi}}{2}\right) \left(L^{B}_{1} + L^{B}_{-1}\right) + \dfrac{l_{u}}{2} \csc \left(\dfrac{l_{\phi}}{2}\right) \left(M^{B}_{1} + M^{B}_{-1}\right) - l_{u} \cot \left(\dfrac{l_{\phi}}{2}\right) M^{B}_{0} - 2 L^{B}_{0} \biggr) \\[0.7ex]
				{}&= - \pi \csc \left(\dfrac{l_{\phi}}{2}\right) \left(l_{u} \cos \left(\phi\right) \csc \left(\dfrac{l_{\phi}}{2}\right) + 2 u \sin \left(\phi\right) \cos \left(\dfrac{l_{\phi}}{2}\right) - l_{u} \cot \left(\dfrac{l_{\phi}}{2}\right)\right) \partial_{u} \\[0.7ex]
				{}& \quad + \pi \csc \left(\dfrac{l_{\phi}}{2}\right) \left(l_{u} \cos \left(\phi\right) \csc \left(\dfrac{l_{\phi}}{2}\right) + 2(r + u) \sin \left(\phi\right) \cos \left(\dfrac{l_{\phi}}{2}\right)\right) \partial_{r} \\[0.7ex]
				{}& \quad + \pi \csc \left(\dfrac{l_{\phi}}{2}\right) \biggl( 2 \cos \left(\dfrac{l_{\phi}}{2}\right) \cos \left(\phi\right) + \dfrac{1}{r} \left(-l_{u} \sin \left(\phi\right) \csc \left(\dfrac{l_{\phi}}{2}\right) + 2 u \cos \left(\phi\right) \cos \left(\dfrac{l_{\phi}}{2}\right)\right) - 2 \biggr) \partial_{\phi}.
			\end{split}
			\label{23}
		\end{equation}
\end{widetext}

The timelike extremal curves can be obtained from the fixed points of the bulk flow $\zeta_{tt}^{B}$ as follows:

\begin{equation}
	\begin{split}
	r \sin \left(\phi\right) \sin \left(\dfrac{l_{\phi}}{2}\right) = - \dfrac{l_{u}}{2}, \\
	r + u  = r  \cos \left(\phi\right) \sec \left(\dfrac{l_{\phi}}{2}\right).
	\end{split}
	\label{55}
\end{equation}
As shown in Fig.~\ref{fig:3}, the timelike extremal curve (depicted in red) is intersected at $r=r^f_I$ by the two null rays emanating from the different endpoints at null infinity, namely

\begin{equation}
	(u,\phi)=\left(0,\frac{l_{\phi}}{2}\right),
	\qquad
	(u,\phi)=\left(0,-\frac{l_{\phi}}{2}\right).
\end{equation}
The coordinates of the two intersection points are
\begin{equation}
	\begin{aligned}
		\beta_{1} =	(u_1,\phi_1,r_I^{f}) &=
		\left(
		-l_u\csc^2\dfrac{l_\phi}{2},
		\dfrac{l_\phi}{2}-\pi,
		r_I^{f}
		\right),\\
		\beta_{2} =	(u_2,\phi_2,r_I^{f}) &=
		\left(
		0,
		-\dfrac{l_\phi}{2},
		r_I^{f}
		\right).
	\end{aligned}
\end{equation}

The modular flows \eqref{78} and \eqref{23} constructed here are the flat space limits of the AdS modular flows \eqref{50} and \eqref{48}, respectively. Similarly, the flat space limit of the AdS timelike extremal curve \eqref{18} gives the flat space timelike extremal curve \eqref{55}, as shown in \cite{Fareghbal_2026}

Using the transformations \eqref{56} and \eqref{57}, the modular flow \eqref{23} takes a simple form in Cartesian coordinates,
\begin{equation}
	\zeta^{B}_{tt} = 2 \pi \left(x \partial_{y} - y \partial_{x}\right),
\end{equation}
which is a rotation in the $(x,y)$ plane.

Substituting the extremal curve \eqref{55} into the metric \eqref{12}, the induced line element becomes
\begin{equation}
	ds^{f}_{\zeta_{tt}} = -\dfrac{2  r \sin^{2}  \left(\dfrac{l_{\phi}}{2}\right)  \sec \left(\dfrac{l_{\phi}}{2}\right)}{\sqrt{l_{u}^2-2 r^2 \left(1 - \cos \left(l_{\phi}\right)\right) }} \: dr .
\end{equation}
Following the calculation of the real part, we assume that only the segment of the timelike curve between the two null rays contributes to the holographic timelike entanglement entropy, and we evaluate its length over $r_{min}^{flat} < r < r_{I}^{flat}$. Therefore,

\begin{equation}
	L_{tt}^{f} = 2 \int_{r_{\min}^{f}}^{r_{I}^{f}} ds^{f}_{\zeta_{tt}} = i l_{u} \csc \left(\dfrac{l_{\phi}}{2}\right).
	\label{30}
\end{equation}
Combining the real and imaginary parts, the total length associated with the timelike interval reads
\begin{equation}
	L^{f}_{tot} = l_{u} \cot \left(\dfrac{l_{\phi}}{2}\right) + i l_{u} \csc \left(\dfrac{l_{\phi}}{2}\right).
	\label{52}
\end{equation}
The holographic timelike entanglement entropy is then
\begin{equation}
	S^{(T)}_{flat} = \dfrac{L^{f}_{tot} }{4 G_{N}} = \dfrac{l_{u}}{4 G_{N}}  \cot \left(\dfrac{l_{\phi}}{2}\right) +  \dfrac{ il_{u}}{4 G_{N}}  \csc \left(\dfrac{l_{\phi}}{2}\right).
	\label{62}
\end{equation}
This quantity is complex, as in AdS$_3$. Unlike the AdS$_3$
result, the imaginary part here is not constant but depends on both  $l_u$ and $l_\phi$. Our result for the imaginary part differs from the one in
\cite{Fareghbal_2026}, where the imaginary part equals the real part,
$l_u\cot(l_\phi/2)$. The difference comes from the value of $\Delta$
at $r=r_I$ on the timelike curve: in \cite{Fareghbal_2026} the value of
$\Delta$ at $r_I$ computed on the spacelike curve is used, whereas the
timelike curve has different $(u,\phi)$ at the same $r$, and hence a
different $\Delta$.

\subsection{Flat Limit of the Holographic Timelike Entanglement Entropy}

We now ask whether the lengths in AdS have a flat space limit that reproduces the corresponding flat space lengths. From \eqref{53} and \eqref{54}, we see that the AdS lengths do not. Following \cite{Fareghbal:2024lqa}, we avoid this problem by keeping only the portion of the lengths between  $r_{\min}$ and $r_{I}$, instead of the full lengths.

The coordinate $r_{I}$ is the radius at which $u$ attains its extremum along the spacelike extremal curve. In AdS$_3$ it is determined by combining the equations in \eqref{16}, which give
\begin{equation}
	\begin{multlined}
		r \sin \left(\dfrac{u}{\ell}\right) - \ell \cos \left(\dfrac{u}{\ell}\right) = 
		\sin \left(\dfrac{l_{\phi}}{2}\right) \sqrt{r^{2} + \dfrac{\ell^{2} \cos^{2} \left(\dfrac{l_{u}}{2 \ell}\right)}{\sin^{2} \left(\dfrac{l_{\phi}}{2}\right) - \sin^{2} \left(\dfrac{l_{u}}{2 \ell}\right)}}
	\end{multlined}
\end{equation}
Setting $du/dr=0$ in this relation yields
\begin{equation}
	r_{I} = \dfrac{\ell \sin \left(\dfrac{l_{u}}{2 \ell}\right) \cos \left(\dfrac{l_{u}}{2 \ell}\right)}{\sin^{2} \left(\dfrac{l_{\phi}}{2}\right) - \sin^{2} \left(\dfrac{l_{u}}{2 \ell}\right)}.
\end{equation}
In the flat space limit $r_I$ reduces to $r_I^f$ given in \eqref{79}.

Integrating the line elements \eqref{32} and \eqref{33} from $r_{\min}$ to $r_I$, we find

\begin{widetext}
	\begin{footnotesize}
		\begin{align}
			L &= 2 \int_{r_{\min}}^{r_{I}} ds_{st} + 2 \int_{r_{\min}}^{r_{I}} ds_{tt} \nonumber = \ell \log \left(\dfrac{1 - \cos \left(l_{\phi} + \dfrac{l_{u}}{\ell}\right)}{\cos \left(\dfrac{l_{u}}{\ell}\right) - \cos \left(l_{\phi}\right)}\right) \nonumber \\[0.7ex]
			& \quad - \ell \log \left( 1 + \dfrac{8 \cos^{2} \left(\dfrac{l_{\phi}}{2}\right) \sin^{2}\left(\dfrac{l_{u}}{2 \ell}\right) + 4 \cos \left(\dfrac{l_{\phi}}{2}\right) \sin \left(\dfrac{l_{u}}{2 \ell}\right) \sqrt{4 \cos^{2}\left(\dfrac{l_{\phi}}{2}\right) \sin^{2} \left(\dfrac{l_{u}}{2 \ell}\right) + \cos^{2} \left(l_{\phi}\right) - \cos^{2} \left(\dfrac{l_{u}}{\ell}\right)}}{\cos^{2} \left(l_{\phi}\right) - \cos^{2} \left(\dfrac{l_{u}}{\ell}\right)}\right).
		\end{align}
	\end{footnotesize}
\end{widetext}
In the flat space limit, this expression reduces precisely to the flat space result \eqref{52} and thus holographically reproduces  CCFT$_2$ timelike entanglement entropy \eqref{62}.

\section{Conclusion}
\label{Conclusion}

	In this work we constructed modular flows for spacelike and timelike intervals in AdS$_3$/CFT$_2$, and extended the same construction to Flat$_3$/CCFT$_2$. Starting from the globally well-defined symmetry generators of the vacuum states of the boundary theories defined on a cylinder, we derived the corresponding boundary modular flows and extended them into the bulk via the dual Killing vector fields of the corresponding gravity theories. In both dualities, the fixed points of these bulk flows coincide with the extremal curves associated with the intervals. For these symmetric configurations, this gives a geometric route to the (timelike) holographic entanglement entropy that does not require solving the extremization problem.

For spacelike intervals, this construction reproduces the known result in AdS$_3$/CFT$_2$. For timelike intervals, the modular flow associated with the timelike extremal curve had not previously been constructed explicitly; here we obtained it directly, together with a second vector field orthogonal to it. The two fields satisfy a fixed norm relation. Under the exchange $l_x \leftrightarrow l_t$, they transform into each other. The curve associated with this second field has no real embedding in the bulk, yet its analytic continuation to complex coordinates gives a length identical to that of the spacelike extremal curve. What this length is actually computing, and why the continuation works so cleanly, is not yet clear to us. A geometric picture of the timelike extremal curve is given in \cite{shalamzari2026holographicpseudoentropyrindler}.

The same construction extends naturally to Flat$_3$/CCFT$_2$, where we obtained the boundary and bulk modular flows associated with timelike intervals. As in AdS$_3$, we look for the extremal curves on which the bulk modular flow vanishes. For a spacelike interval, there is a single extremal curve, but not all of it contributes to the entropy: only the piece cut off between two null rays does \cite{Jiang_2017}. The same thing happens for a timelike interval, except now there are two curves, a spacelike curve and a timelike curve. The spacelike curve works just like before, cut off by null rays at the same pair of points. The timelike curve is cut off too, but by null rays coming from a different pair of points. We found that the contribution from the spacelike curve reproduces the standard CCFT$_2$ entanglement entropy, while the timelike curve gives rise to an additional imaginary contribution. Clarifying the physical interpretation of this term remains an interesting direction for future studies. 

Taking the flat space limit of the full AdS extremal length does not reproduce the flat space result directly, since the AdS length diverges in this limit. Following the prescription of \cite{Fareghbal:2024lqa}, we resolve this by restricting to the portion of the length of the AdS extremal curve, the segment between the minimum radial coordinate $r_{\min}$ and the point $r_I$ where the retarded time $u$ is extremized. The flat limit of this effective length reproduces the same result obtained directly from flat space holography.

In summary, these results provide modular flow as a common geometric formalism for holographic entanglement entropy, applicable to both spacelike and timelike regions in AdS as well as in flat space. Two directions remain open for future work: extending this method beyond three dimensions, and understanding the physical origin of the imaginary contributions.

\begin{acknowledgments}
We appreciate Reza Fareghbal’s helpful comments on an earlier version of this paper. MHSH was supported in part by Shahid Chamran University of Ahvaz under grant no. SCU.SP1404.86612. MKZ thanks Shahid Chamran University of Ahvaz, Iran, for supporting this work under research grant no. SCU.SP1404.37271.
\end{acknowledgments}

	\nocite{REVTEX42Control,apsrev42Control}
\bibliographystyle{apsrev4-2}
\bibliography{refrences1}

\end{document}